\documentclass[%
    superscriptaddress,%
    amsmath,%
    amssymb,%
    aps,
    pra,%
    twocolumn,%
    floatfix,%
]{revtex4-2}

\usepackage{graphicx}
\usepackage{bm,bbm}
\usepackage[colorlinks=true,
bookmarks=false,
linkcolor=blue,
urlcolor=blue,
citecolor=blue,
breaklinks]{hyperref}
\usepackage{xcolor}
\usepackage{physics}
\usepackage{comment}

\usepackage[ruled,vlined]{algorithm2e}

\begin{document}

\title{The Bonsai algorithm: grow your own fermion-to-qubit mappings}

\author{Aaron Miller}
\email{aaron.miller@algorithmiq.fi}
\affiliation{Algorithmiq Ltd, Kanavakatu 3C, FI-00160 Helsinki, Finland}
\affiliation{School of Physics, Trinity College Dublin, College Green, Dublin 2, Ireland}
\author{Zolt\'{a}n Zimbor\'{a}s}
\affiliation{Algorithmiq Ltd, Kanavakatu 3C, FI-00160 Helsinki, Finland}
\affiliation{Wigner Research Centre for Physics, H-1525, P.O.Box 49, Budapest, Hungary}
\author{Stefan Knecht}
\affiliation{Algorithmiq Ltd, Kanavakatu 3C, FI-00160 Helsinki, Finland}
\affiliation{ETH Z{\"u}rich, Laboratory for Physical Chemistry, Vladimir-Prelog-Weg~2, 8093 Z{\"u}rich, Switzerland}
\author{Sabrina Maniscalco}
\affiliation{Algorithmiq Ltd, Kanavakatu 3C, FI-00160 Helsinki, Finland}
\author{Guillermo Garc\'{i}a-P\'{e}rez}
\affiliation{Algorithmiq Ltd, Kanavakatu 3C, FI-00160 Helsinki, Finland}

\date{\today}

\begin{abstract}
    Fermion-to-qubit mappings are used to represent fermionic modes on quantum computers, an essential first step in many quantum algorithms for electronic structure calculations.
    In this work, we present a formalism to design flexible fermion-to-qubit mappings from ternary trees.
    We intuitively discuss the connection between the generating trees' structure and certain properties of the resulting mapping, such as Pauli weight and the delocalisation of mode occupation.
    Moreover, we introduce a recipe that guarantees Fock basis states are mapped to computational basis states in qubit space, a desirable property for many applications in quantum computing.
    Based on this formalism, we introduce the Bonsai algorithm, which takes as input the potentially limited topology of the qubit connectivity of a quantum device and returns a tailored fermion-to-qubit mapping that reduces the SWAP overhead compared to other paradigmatic mappings.
    We illustrate the algorithm by producing mappings for the heavy-hexagon topology widely used in IBM quantum computers.
    The resulting mappings have a favourable Pauli weight scaling $\mathcal{O}(\sqrt{N})$ on this connectivity while ensuring that no SWAP gates are necessary for single excitation operations.
\end{abstract}

\maketitle

\section{Introduction}
\label{sec:intro}
The field of quantum computing has witnessed astounding developments in the last decade.
While the technology is improving quickly, so-called fault-tolerant quantum computing still seems a distant milestone.
Current devices are limited to relatively few qubits and cannot reliably execute the deep circuits required by many paradigmatic quantum computing algorithms \cite{preskill2018quantum}.
Yet, near-term computers with a few hundred qubits and low levels of noise can prepare entangled states that cannot be efficiently simulated classically, which can be a computational resource in itself if combined with classical compute appropriately \cite{arute2019quantum,zhong2020quantum,wu2021strong,madsen2022quantum}, for instance, to mitigate the detrimental effects of noise \cite{endo2018practical,endo2021hybrid,cai2022quantum,berg2022probabilistic}.
This is the reasoning behind most hybrid quantum-classical computing approaches \cite{bharti2022noisy}.
Within this hybrid framework, it was recognised early on that one application stands out from the rest in terms of suitability: the simulation of many-body fermionic quantum systems.

Solving electronic structure problems with near-term devices is of paramount importance in fields such as computational chemistry \cite{kandala2017hardware,barkoutsos2018quantum,ollitrault2020quantum}, which itself has a crucial impact on many industries ranging from material science \cite{lordi2021advances} to drug discovery \cite{cao2018potential,blunt2022perspective}, amongst many others.
In many cases, the limitations of classical methods for chemistry stem from the inability to account for the complexity of electronic wave functions, which can easily involve a superposition of a combinatorially large number of electronic configurations.
Given that quantum processors can physically exhibit complex superpositions, this problem seems particularly appropriate for them to tackle.
Indeed, in most near-term approaches, the physical state of the device is taken to represent the state of the many-body fermionic system of interest, and the physical properties of the latter are inferred by appropriately measuring physical properties of the former \cite{mcardle2020quantum,tilly2022variational}.
However, since the fermionic and the many-qubit wavefunctions live in fundamentally different Hilbert spaces, equipped with different algebraic structures, realising this prospect calls for a concrete way to establish connections between them, the so-called fermion-to-qubit mappings.
The type of mapping between fermions and qubits used has a direct impact on the quantum simulation.
A fermionic wavefunction $\ket{\Psi_\mathrm{f}}$ will be encoded into different many-qubit states $\ket{\Psi_\mathrm{q}}$ by different mappings, and these different states will generally not be equally easy to prepare on a given quantum processor.
Ultimately, the latter point depends on the specifics of the hardware used, e.g., its connectivity, coherence time, etc.
Moreover, one is not generally interested in merely simulating state $\ket{\Psi_\mathrm{f}}$, but in determining physical properties, $\bra{\Psi_\mathrm{f}} \mathcal{O}_\mathrm{f} \ket{\Psi_\mathrm{f}}$, where $\mathcal{O}_\mathrm{f}$ is some fermionic quantity of interest.
The fermion-to-qubit mapping of choice will map said operator into its qubit counterpart, $\mathcal{O}_\mathrm{q}$, and the evaluation of $\bra{\Psi_\mathrm{q}} \mathcal{O}_\mathrm{q} \ket{\Psi_\mathrm{q}}$ will be carried out via physical measurements on the device.
Therefore, in general, the measurement cost incurred will also depend on the choice of mapping.
The importance of the encoding has thus motivated considerable research towards designing convenient fermion-to-qubit mappings beyond the paradigmatic ones, such as Jordan-Wigner (JW) \cite{jordan_uber_1928}, Bravyi-Kitaev (BK) \cite{bravyi2002fermionic}, and Parity \cite{Bravyi2017}.
Much work in relation to the design of fermion-to-qubit mappings is directed toward the reduction of qubit and Pauli weight (the number of qubits that mapped fermionic operations involve) requirements on lattice models \cite{Chiew2022, Setia2018, Steudtner2018, Whitfield, Chien2020, Steudtner2019}. 
Some works avoid fermionic encoding of the wave function altogether in an effort to mitigate the associated costs \cite{mazziotti2021quantum, PhysRevA.105.062424}
Other works have addressed the measurement cost in certain fermionic simulation tasks, which directly depends on the Pauli weight of the mapping when using informationally complete measurements, and thus introduced a mapping with provable optimal Pauli weight \cite{jiang2020optimal}.
The latter mapping is generated using regular ternary trees and largely inspired the current work.
However, the connection between ternary trees and mappings was established earlier.
In Ref.~\cite{vlasov2019clifford}, it was used to find representations of the Clifford algebras and spin groups, and it was discussed that JW in BK can be generated from ternary trees as linear and binary subgraphs of them with the appropriate pairing of the resulting strings.
More recently, a framework to design mappings based on BK was put forward in Ref.~\cite{Li2022}, but it cannot achieve optimal Pauli weight scaling.

In this paper, we consider the family of mappings generated by identifying Majorana operators with linearly and algebraically independent Pauli strings obeying equivalent anti-commutation relations.
In Section~\ref{sec:theory}, we give an overview of fermion-to-qubit mappings and present a classification in terms of the number of qubits that the non-trivial overlap (NTO) between strings involves.
We then focus on one-qubit-wise anti-commuting (1-NTO) encodings and introduce a ternary-tree-based framework to design mappings within this sub-class.
More precisely, we prove that any $n$-node connected ternary tree yields a valid mapping.

Importantly, this framework presents a number of desirable properties.
On the one hand, it is clear and intuitive.
It enables understanding, and consequentially the control of many important properties of the mappings, such as their mode locality (that is, on how many qubits fermionic mode occupation is stored, which has been suggested to impact the resilience of quantum simulations to noise \cite{sawaya2016error}) and their Pauli weight, in terms of simple properties of the underlying ternary trees generating them.
Part of the discussion in Section \ref{sec:theory} is aimed at explaining these aspects in a pedagogical manner.
On the other hand, it contains the aforementioned paradigmatic mappings, as well as the recently discovered optimal-weight fermion-to-qubit mapping \cite{jiang2020optimal}, as specific instances.
Hence, we can regard our framework as being able to \textit{interpolate between} and \textit{combine} well-known mappings, as well as generate completely novel ones that bear little resemblance with these.
In addition, these two properties together provide, as a byproduct, a clean and transparent perspective on these widely used encodings and their properties.

We also show that with the right assignment between Majorana operators and Pauli strings, for which we provide a recipe, the tree-based mapping design framework introduced here guarantees a crucial property of the sampled encodings: uncorrelated fermionic states, i.e., Fock basis states (including the vacuum state), are mapped to computational basis states in qubit space.
This property, which may be easily overlooked, is of utmost importance, as it ensures no additional quantum resources (more precisely, entanglement) are needed to prepare reference wavefunctions, such as the Hartree-Fock Slater determinant often used as a starting point in many quantum simulation algorithms.

In Section \ref{sec:hardware_efficient_mappings}, we put our framework into use.
We exploit its versatility to devise an algorithm, which we name \textit{the Bonsai algorithm}, that takes a hardware connectivity layout as an input and returns a tailored mapping.
The resulting encoding is designed to minimise SWAP overhead required in the implementation of one- and two-electron excitation operations, which are the building blocks of many adaptive ansatz construction algorithms of the ADAPT-VQE family \cite{tang2021qubit,grimsley2019adaptive,anastasiou2022tetris}.
At the same time, the algorithm aims at minimising the spread of fermionic occupancy over qubits.
When applied to heavy-hexagon qubit connectivity graphs, the standard layout in current IBM devices, the Bonsai algorithm returns a mapping with no SWAP overhead for single-excitation operations and a quadratically lower Pauli weight than JW.
Moreover, the mapping also drastically reduces circuit complexity of the worst-case scenario implementation of single- and double-excitation generated unitaries with respect to the latter mapping.

\section{Fermion-to-qubit mappings}
\label{sec:theory}


This section is devoted to the discussion of features of general fermion-to-qubit mappings.  After setting up the notation for fermionic systems (we refer the reader to Refs.~\cite{alickifannes, zimboras2014dynamic, szalay2021fermionic} for detailed reviews),  the Jordan-Wigner transformation, general  fermion-to-qubit maps, and  finally Majorana string mappings are described.

\subsection{Fermionic systems}
Consider an $N$-mode fermionic system in second quantisation described in terms of $N$ creation and annihilation operators, $\{ a^\dagger_i \}_{i = 0, \ldots, N-1}$ and $\{ a_i \}_{i = 0, \ldots, N-1}$, which fulfil the usual canonical fermionic anti-commutation relations 
\begin{align}
   \{ a_i, a_j \} = \{ a^\dagger_i, a^\dagger_j \} = 0 \quad \text{ and } \quad \{ a^\dagger_i, a_j\} = \delta_{ij} \mathbbm{1}
\end{align}
The creation and annihilation operators act on $\mathcal{F}(\mathbb{C}^N)$, the Fock space belonging to an $N$-dimensional one-particle space. This is a $2^N$ dimensional Hilbert space spanned by the fermionic vacuum $\ket{\text{vac}_\text{f}}$ and the vectors obtained by applying subsets of fermionic creation operators; this orthonormal basis, also called Fock basis, can be denoted as
\begin{align}
    \ket{n_0, n_1, ... \, n_{N{-}1}}{:=} (a^\dagger_0)^{n_0} (a^\dagger_1)^{n_1} ...\, (a^\dagger_{N{-}1})^{n_{{N{-}1}}} \ket{\text{vac}_\text{f}}, 
\end{align}
where $n_j \in \{0,1 \}$ are the so-called occupation numbers of mode $j$ and the notation $(a^\dagger_j)^{0}=\mathbbm{1}$ is used.
The fermion-number operator for mode $j$ is given as $\hat{n}_j=a^\dagger_j a_j$.
It is easy to show that the Fock basis states are eigenstates of these local fermion-number operators with eigenvalue given by the occupation numbers, e.g., $\hat{n}_0 \ket{n_0, n_1, \ldots n_{N-1}}= n_0  \ket{n_0, n_1, \ldots n_{N-1}} $.

Besides the fermionic creation and annihilation operators, another useful set of generators for the fermion observables are the $2N$ Majorana operators $\{ m_k \}_{k = 0, \ldots, 2N-1}$ defined as
\begin{equation}
    \label{eq:fermion_to_Majorana}
    m_{2 j}=a_{j}^{\dagger}+a_{j} \quad \text{and} \quad m_{2 j+1}=i\left(a_{j}^{\dagger}-a_{j}\right),
\end{equation}
which are unitary, self-adjoint, and obey the Majorana anticommutation relations
\begin{equation}
    \label{eq:Majorana_antcomm}
    \{m_{i},m_{j}\}=2\delta_{ij}\mathbbm{1}.
\end{equation}
Any fermionic observable can be uniquely expressed as a linear combination of Majorana monomials $m_{x_0}...\,m_{x_j}$. In particular, the number operator for mode $j$ is $\hat{n}_j=\frac{1}{2}(\mathbbm{1}+im_{2j}m_{2j +1})$. 

\subsection{Jordan-Wigner transformation and general fermion-to-qubit mappings}

The fermionic Fock space $\mathcal{F}(\mathbb{C}^N)$ and the Hilbert space of $N$ qubits $(\mathbb{C}^{2})^{\otimes N}$ are both $2^{N}$ dimensional Hilbert spaces, thus one can map one into the other unitarily.
A very natural unitary mapping is to map the Fock basis states of $\mathcal{F}(\mathbb{C}^N)$ to the computational basis states of the qubits such that the occupation number of the $j$-th fermionic mode matches with the state of the $j$-th qubit \cite{jordan_uber_1928}:
\begin{equation}
  \mathcal{F}(\mathbb{C}^N)  {\ni}  \ket{n_0, n_1, \ldots ,n_{N{-}1}} \mapsto \bigotimes \limits_{i = 0}^{N - 1} \ket{n_i} {\in} (\mathbb{C}^{2})^{\otimes N}.
\end{equation}
On the operator level, this correspondence induces a linear mapping between the corresponding observable algebras given by:
\begin{align}
 m_{2j} \mapsto X_j\prod_{k=0}^{j-1} Z_k, \\
 m_{2j+1} \mapsto  Y_j \prod_{k=0}^{j-1} Z_k, 
\end{align}
for $j=0,1,\ldots N-1$. Here and in the rest of the paper, we use the notation $P_j$ with $P \in \{X,Y,Z\}$ for an operator that acts as the Pauli operator $P$ on the $j$-th qubit and as identity on the other qubits. 

In general, a unitary mapping between the fermionic and qubit Hilbert spaces induces a linear mapping on the corresponding observable algebras such that
\begin{align}
    & m_{k} \mapsto R_k, \quad k=0, \ldots 2N-1, \label{eq:GenF2Q}\\
    &\text{(i) $R_k$'s are algebraically independent} ,\nonumber \\
    &\text{(ii)}\;\{R_k, R_\ell \}= 2\delta_{k\ell}\mathbbm{1}. \nonumber
\end{align}

Conversely, any linear mapping between the fermionic and qubit observable algebras satisfying the properties (i) and (ii) in Eq.\eqref{eq:GenF2Q} defines uniquely (up to a global phase factor) a unitary mapping between $\mathcal{F}(\mathbb{C}^N)$ and the Hilbert space of $N$ qubits $(\mathbb{C}^{2})^{\otimes N}$. 
This unitary mapping between the two Hilbert spaces can be constructed as follows: Since  the fermionic vacuum state  $\ket{\text{vac}_\text{f}}$ is the unique vector (up to a scalar factor) satisfying the relations $a_j \ket{\text{vac}_\text{f}} =\tfrac{1}{2}(m_{2j} + i m_{2j +  1}) \ket{\text{vac}_\text{f}}= 0 $ for all $j=0, \ldots, N-1$, the vacuum state is mapped to the state  $\ket{\psi}$ which satisfies that $\tfrac{1}{2} (R_{2j} + i R_{2j +  1}) \ket{\psi} = 0$ for all $j=0, \ldots, N-1$ (note that such a $\ket{\psi}$ is unique up to a phase factor). Any other Fock basis vector  $ a^\dagger_{j_0} a^\dagger_{j_1} \ldots a^\dagger_{j_\ell} \ket{\text{vac}_\text{f}}$ is mapped to $ \tfrac{1}{2} (R_{2j_0} {-} i R_{2j_0 +  1}) \tfrac{1}{2} (R_{2j_1} {-} i R_{2j_1 +  1})  ... \tfrac{1}{2} (R_{2j_\ell} {-} i R_{2j_\ell +  1}) \ket{\psi}$.

\subsection{Majorana string mappings}\label{subsec:MString}

When it comes to mapping fermionic systems to qubit systems, the Pauli basis suggests a path: finding a suitable set $\mathcal{S}$ of $2N$ Pauli strings (i.e., products of Pauli operators) $S_k$ ($k=0 \ldots 2N-1$) fulfilling the anti-commutation $\{ S_{i}, S_{j} \} = 2 \delta_{ij} \mathbbm{1}$.
For this approach to result in a proper fermion-to-qubit mapping, the Pauli strings in $\mathcal{S}$ must also be linearly and algebraically independent.
Linear independence is trivially satisfied if all the strings differ, but algebraic independence is more subtle.
This means that it must not be possible to find two different subsets $A \subseteq \mathcal{S}$ and $B \subseteq \mathcal{S}$, $A \neq B$, such that $\prod_{S_i \in A} S_i \propto \prod_{S_j \in B} S_j$, given that the corresponding products of Majorana operators in fermion space result in distinct operators.
Throughout this work, we will use the term \textit{Majorana strings} to refer to the Pauli strings within a set $\mathcal{S}$ satisfying these conditions.  

Summarising the above, in this paper we will consider so-called {\it Majorana string fermion-to-qubit mappings}, which are linear mappings between the fermionic and qubit system observable algebras that satisfy the following criteria:
\begin{itemize}
    \item {\bf Criterion~(A)}: Each Majorana operator is mapped to a  Pauli string,  
    $m_j \to  S_j \in \mathcal{S}$ for $j=0, \ldots, 2N-1$. 
    \item {\bf Criterion~(B)}:  The above Pauli strings satisfy $ \{ S_{i}, S_{j} \} = 2 \delta_{ij}\mathbbm{1}$. 
    \item {\bf Criterion~(C)}: For any unequal subsets $A \subseteq \mathcal{S}$ and $B \subseteq \mathcal{S}$, $A \neq B$, $\prod_{S_i \in A} S_i \propto \prod_{S_j \in B} S_j$ is not fulfilled. 
\end{itemize}

Furthermore, one often considers Majorana string fermion-to-qubit mappings that satisfy an additional criterion:
\begin{itemize}
    \item {\bf Criterion~(D)}: Vacuum preservation, the fermionic vacuum is mapped to the all-zero computational basis state, i.e.,  $\ket{\text{vac}_\text{f}} \mapsto \ket{0}^{\otimes N}$.
\end{itemize}
Mappings  satisfying also  Criterion (D) besides Criteria (A)-(C), will be called {\it vacuum preserving} Majorana string fermion-to-qubit mappings.

Consider a Majorana string $S_j$ which is (up to a phase factor) a product of (non-identity) Pauli operators over a subset of sites $A_j \subset \{0, \ldots, N-1 \}$.
We call $A_j$ the {\it support of  $S_j$}. Let $A_j$ and $A_k$ be the supports of $S_j$ and  $S_k$, respectively.
We call  $A_j \cap A_k$ the {\it overlapping sites} of $S_j$ and  $S_k$.
The subset of  $N_{j,k} \subseteq A_j \cap A_k$ where the local Paulis corresponding of the Majorana strings $S_j$ and  $S_k$ are different is called the {\it non-trivial overlapping sites of $S_j$ and $S_k$}.
As any two Majorana strings anticommute, the number of non-trivial overlapping sites of any pair of Majorana strings must be odd.
Given a Majorana string fermion-to-qubit mapping, let $k$ be the maximum of this odd number considering all the pairs of Majorana strings.
We call such a mapping a $k$-non-trivial-overlap ($k$-NTO) Majorana string fermion-to-qubit mapping.
Most of the known fermion-to-qubit mappings (e.g., JW, BK, and Parity) are 1-NTO, but non-1-NTO mappings also exist (see Appendix \ref{app:exotic}).

\section{Mappings originating from general ternary trees}
\label{sec:mappings_from_trees}
In this subsection, we explore the correspondence between a certain class of graphs, called ternary trees (TT), and vacuum-preserving fermion-to-qubit mappings.
As mentioned previously, the connection between the two has been established before \cite{vlasov2019clifford, jiang2020optimal}.
In Ref.~\cite{jiang2020optimal}, the minimum-depth TT is used to find a fermion-to-qubit mapping with optimal Pauli weight.
Inspired by that work, we now extend the TT formalism and show that \textit{any} TT can result in a valid mapping, the properties of which can be directly connected to the graph-theoretical properties of the tree.
This connection will be explored more carefully in the next subsection.
Moreover, the mapping-generating method introduced here guarantees that, for any sampled mapping, the fermionic vacuum is mapped to the all-zeros qubit state, and Fock basis states are mapped to computational basis states; we refer to this property as \textit{product preservation}.

\subsection{Ternary trees}
It will be useful for this section and the next one to start by reviewing some graph-theoretic concepts.
A graph is a pair $\mathcal{G} = (V,E)$ where $V$ is a set of vertices or nodes, and $E \subseteq\left\{(x, y) \mid (x,y) \in V^2, x \neq y\right\}$ are the edges, or links, which are sets of paired vertices.
We consider undirected graphs, meaning that $(x, y) \in E \Rightarrow (y, x) \in E$.
A length-$\ell$ path $\mathbf{p}$ in a graph is a length-$\ell$ ordered sequence of vertices $\mathbf{p} = \{ p_0, p_1, \ldots, p_{\ell - 1} \}$ such that any pair of consecutive vertices in the sequence are connected in the graph, that is, for every $l < \ell$, $(p_{l-1}, p_{l}) \in E$.
We call a graph $\mathcal{G}$ connected if, for any pair of vertices $u, v$, there exists at least one path $\mathbf{p}$ with $u$ and $v$ as endpoints.
In any such graph $\mathcal{G}$, the path structure induces a well-defined metric distance $d(u, v)$ between all pairs of vertices, defined as the length $\ell$ of the shortest path (or paths, as they may not be unique) with the two vertices as endpoints.
It can be easily shown that the set of distances $d(u, v)$ define a proper metric space, given that they are positive, symmetric, zero iff $u = v$, and they fulfil the triangle inequality, $d(u, v) \leq d(u, w) + d(w, v), \, \forall u, v, w \in V$.

The path structure also allows us to define a special kind of graph, the tree.
A tree $\mathcal{T}$ is a graph that contains no loops, that is, for which there are no paths $\mathbf{p}$ containing any node more than once.
It is a fact that any connected $N$-node tree contains exactly $N - 1$ edges, and that any connected undirected graph with $N - 1$ edges is a tree.
It is also useful to define the degree of a node $u$ as the number of edges reaching $u$,
\begin{equation}
    \label{eq:degree}
    \Delta(u) = \vert \{v \in V \mid (u, v) \in E\} \vert.
\end{equation}

\begin{figure}[t]
   \centering
   \includegraphics[width=\columnwidth]{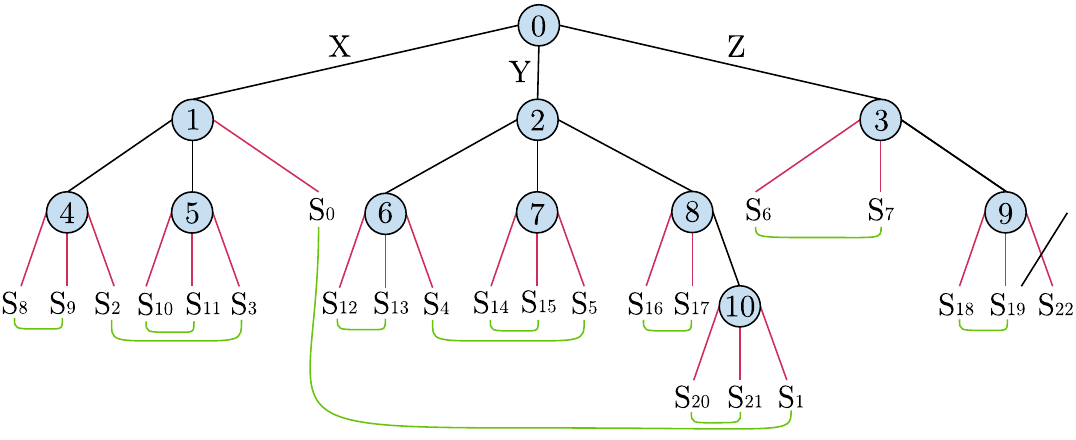}
   \caption{Example of a mapping derived from a ternary tree.
   The black lines represent the edges, which always connect two qubits, while the legs are depicted in red.
   Throughout this paper, we always represent the Pauli labelling on the links (both edges and legs) by their position: the left-most link below a node is labelled with an $X$, while the right-most with a $Z$.
   Every leg in the tree can be associated with a Pauli string by following the path from the root node (0, in this case) to the leg.
   Every time a link with label $P$ stemming downwards from a qubit $u$ is crossed, the operator $P$ acting on qubit $u$ is added to the string.
   The resulting string acts trivially on all qubits not visited along the path.
   For example, with the tree in the figure, the leg labelled with $S_{0}$ generates the string $S_0 = X_0Z_1$, while $S_1 = Y_0 Z_2 Z_8 Z_{10}$.
   The green lines depict the pairing between Majorana strings that guarantees product preservation.
   Notice that, following the upward paths starting from the two legs of any such pair, the two paths meet at a qubit $u$ once they both cross their first link not labelled with a $Z$.
   In the previous example with strings $S_0$ and $S_1$, this corresponds to the root node 0.}
   \label{fig:example_tree}
\end{figure}

With these definitions at hand, we can introduce the TT.
Essentially, a TT is a tree in which the branching rate is at most three, that is, each node has at most three \textit{descendants}.
To explain this concept more precisely, let us describe a process in which a TT is built by adding nodes sequentially.
First, we start with a single node, the \textit{root} $r$.
Next, we add $k_r \leq 3$ nodes to the graph and connect each of them to the root $r$.
The root now has degree $\Delta(r) = k_r$.
Next, for each of these new nodes (\textit{descendants} of $r$) $u$, we add $0 \leq k_u \leq 3$ nodes, which connect to $u$.
The process can be iterated until the graph contains $N$ nodes.
Notice that, since each time we add a node we add one link along with it, the total number of edges will be $N-1$ (this observation also shows why the resulting graph is a tree: it is not possible to close any loops in the graph by connecting one new node to a single existing one only).
Any node without descendants will be denoted \textit{leaf} throughout this work.

\subsection{From ternary trees to fermion-to-qubit mappings}
The starting point of this work is the observation that, following a simple procedure introduced in Refs.~\cite{vlasov2019clifford,jiang2020optimal} for specific TTs, any $N$-node TT can be used to generate valid fermion-to-qubit mappings.
We now explain this procedure and show its generality.
The scheme is illustrated in Fig.~\ref{fig:example_tree} where we present an 11-qubit example.

Suppose we have an $N$-qubit system that we want to use to simulate an $N$-mode fermionic system.
Without loss of generality, in what follows we label each qubit  with an integer number $u = 0, \ldots, N - 1$.
To generate a fermion-to-qubit mapping, we first generate an $N$-node TT, by for instance following the iterative procedure introduced previously.
Next, we assign a qubit label $u$ to each node.
Different assignments will lead to different mappings (a degree of freedom that can be exploited), but any such labelling is admissible.
In the example (Fig.~\ref{fig:example_tree}), this corresponds to the tree with blue nodes and black edges.

The next step in the process is to add $3 - k_u$ \textit{legs}, that is, edges without a node at the other end of the link, to each node $u$, where $k_u$ is the number of descendants of $u$.
Note that $k_r = \Delta(r)$ for the root $r$, while $k_u = \Delta(u) - 1$ for all other nodes as $\Delta(r) \leq 3$ and $ \Delta(u) \leq 4$ .
These are the red links in Fig.~\ref{fig:example_tree}.
Importantly, doing this will result in $2N + 1$ legs for any tree.
To see this, let us denote the number of legs by $L$. 
Every node except the root is now reached by four \textit{links} (a term that we will use in this work to refer to both edges in the original tree and legs), while the root is reached by three.
Thus, if we sum all the new degrees (including legs) for all the nodes, we obtain $F = 4N - 1$.
In this sum, we have counted each of the edges in the original tree twice (once per each node at its endpoints), while the legs have been counted only once.
Hence, we have $F = |E| + L$, with $|E| = 2(N - 1)$ (since each edge is contained twice in $E$), so $L = 2N + 1$.

Once the legs have been added, and every node has exactly three descending links (either edges or legs), we distribute the labels $X$, $Y$, and $Z$ among said three links of each node.
In order to ease the graphical depictions of the trees in this work, the left-most link implicitly carries the label $X$, the central $Y$, and the right-most one, $Z$, as in Ref.~\cite{jiang2020optimal}.

With this labelling of nodes, edges, and legs, the tree can be used to generate Pauli strings in the following manner.
For every leg, there exists a unique path leading from it to the root $r$.
The path only includes one leg, the starting one, and it may cross some edges as well.
To every link in the path, we can associate a unique Pauli matrix, $P_u$, where $u$ is the parent node reaching the link, and $P$ the label $X$, $Y$, or $Z$, corresponding to the link.
The Pauli string is therefore formed by taking the tensor product of these Pauli operators, along with identity on nodes not along the path.
Since every path results in a different Pauli string, this procedure generates $2N + 1$ strings for $N$ qubits.

Importantly, all these Pauli strings anticommute with one another. 
This can be seen by considering two Pauli strings $S_i$ and $S_j$ stemming from two different legs in the tree.
The two paths corresponding to the legs must meet, that is, when traversing them upwards towards the root, they must have a first node in common (which may be the root itself).
If the first common ancestor to both legs is not the root, both paths from that node upwards are equal, and hence both $S_i$ and $S_j$ contain the same Pauli operators for those qubits.
The first common ancestor, on the other hand, must be reached following two different descendants of said node, so the Pauli matrices for that qubit in the strings are distinct, and both different from identity (what we refer to as \textit{non-trivial overlap}).
If the legs are not directly connected to their common ancestor, their paths include other nodes laying below the latter in the tree.
However, notice that, by definition of a first common ancestor, those nodes cannot be present in both paths, and hence one of the strings must act trivially on each such qubit.
In short, $S_i$ and $S_j$ have a single-qubit non-trivial overlap, that is, for every qubit different from the aforementioned first common ancestor, the corresponding Pauli matrices are either both equal or at least one of them is equal to the identity.

Since all the $2N + 1$ resulting Pauli strings are different, they are obviously linearly independent, but they are not algebraically independent.
In particular, any two disjoint subsets of strings $A$ and $B$, $A \cap B = \emptyset$ such that $A \cup B$ is the whole set of strings fulfil $\prod_{S_i \in A} S_i \propto \prod_{S_j \in B} S_j$.
However, as we prove in Appendix \ref{app:alg_ind_proof}, any subset missing at least one Pauli string is algebraically independent.
Therefore, by dropping any of the strings, the remaining $2N$ ones can be readily identified with the $2N$ Majorana operators associated with the $N$-mode fermionic system, thus defining a valid fermion-to-qubit mapping.

\subsection{Majorana string pairing for product preservation}\label{sec:pairing}
The previous discussion illustrates how any $N$-node TT can be used to obtain $N$-mode mappings.
While any association between the generated Pauli strings and Majorana operators results in a legitimate mapping, not all are equally useful in practice.
Many applications of fermion-to-qubit mappings require that at least some reference state (e.g., the vacuum) be known in qubit space.
In near-term quantum computing, for instance, it is also desirable that Fock basis states be mapped to computational basis states.
We now provide a simple recipe to enforce this product preservation feature of the map, including guaranteeing that the fermionic vacuum is mapped to the state $\ket{0}^{\otimes N}$.

First, let us introduce the concept of \textit{pairing}.
According to Eq.~\eqref{eq:fermion_to_Majorana}, there are two Majorana operators, $m_{2j}$ and $m_{2j + 1}$, associated to every fermionic mode $j$.
Therefore, after identifying Majorana strings with Majorana operators, every creation and annihilation operator $a^{(\dagger)}_j$ will be associated with two Pauli strings.
The key to product preservation lies in how the Pauli strings are paired into fermionic modes.

Let the set of Majorana strings be the set obtained by removing the Pauli string corresponding to the path that only involves links with label $Z$, and consider the following pairing algorithm.
For every node $u$ in the TT, follow its downward link labelled with $X$.
If the link is not a leg, keep travelling downwards taking always the $Z$-links until a leg is reached.
Denote the leg by $s_x^{(u)}$.
The same procedure, starting from the link with label $Y$ will lead to a different final leg $s_y^{(u)}$.
The two Pauli strings $S_{s_x^{(u)}}$ and $S_{s_y^{(u)}}$ should then be paired together into some fermionic mode $j$, that is, one of them should be identified with $m_{2j}$ and the other one with $m_{2j + 1}$.
These pairings are illustrated with green lines in Fig.~\ref{fig:example_tree}.
The simplest identification corresponds to the mapping
\begin{equation}
    \label{eq:pairing_identification}
    m_{2j} \leftrightarrow S_{s_x^{(u)}} \quad \text{and} \quad m_{2j + 1} \leftrightarrow S_{s_y^{(u)}},
\end{equation}
but it is worth mentioning that it is also possible to ensure that the mapped creation and annihilation operators are real in qubit space by associating with $m_{2j}$ the Pauli string that contains an even number of $Y$ operators and with $m_{2j + 1}$ the one containing an odd number of them.
However, for the sake of simplicity, we will only consider the first type of identification explicitly throughout this work.
Importantly, notice that the identification between modes $j$ and qubits $u$ in Eq.~\eqref{eq:pairing_identification} implicitly establishes a bijection between the two sets.

\LinesNumbered
\begin{algorithm}[t]
    \label{alg:pairing_scheme}

    \caption{Pairing scheme}
    
        Choose a bijection $f$ between modes $j$ and qubits $u$, $j = f(u)$.

        Let $V$ be the set of qubits in the tree.
        
        \For{$u \in V$}{
        
            Define $s$ as the $X$-labelled downward link stemming from $u$.
            
            \While{$s$ not a leg}{
            
                Define $v$ as the qubit reached following $s$ downwards.
                
                Define $s$ as the $Z$-labelled downward link stemming from $v$.
                
            }
            
            Set $s \rightarrow s_x^{(u)}$.
            
            Define $s$ as the $Y$-labelled downward link stemming from $u$.
            
            \While{$s$ not a leg}{
            
                Define $v$ as the qubit reached following $s$ downwards.
                
                Define $s$ as the $Z$-labelled downward link stemming from $v$.
                
            }
            
            Set $s \rightarrow s_y^{(u)}$.
            
        }
        
        Remove the unpaired right-most $Z$-leg from the tree.   
        
        Create mode operators $a_j$ and $a^{\dagger}_j$ using Eq.~\eqref{eq:mapped_creation_annihilation} with $j = f(u)$.
        
        Return mapped mode operators.
        
\end{algorithm}

By inverting Eq.~\eqref{eq:fermion_to_Majorana}, we see that the fermionic creation and annihilation operators are mapped into qubit space according to
\begin{equation}
    \label{eq:mapped_creation_annihilation}
    \begin{aligned}
    a^{\dagger}_j &= \frac{1}{2} (m_{2j} - i m_{2j + 1}) \leftrightarrow \frac{1}{2} \left(S_{s_x^{(u)}} - i S_{s_y^{(u)}}\right), \\
    a_j &= \frac{1}{2} (m_{2j} + i m_{2j + 1}) \leftrightarrow \frac{1}{2} \left(S_{s_x^{(u)}} + i S_{s_y^{(u)}}\right).
    \end{aligned}
\end{equation}
From these expressions, it is possible to show that the fermionic vacuum is represented by the qubit state $\ket{0}^{\otimes N}$.
To do so, consider the two Pauli strings $S'_{s_x^{(u)}}$ and $S'_{s_y^{(u)}}$ obtained by substituting the $Z$ operators on the qubits that lie below $u$ on the tree (if any) with identity.
Clearly, $S'_{s_x^{(u)}}$ and $S'_{s_y^{(u)}}$ are equal except for the Pauli operator acting on qubit $u$, which implies that $S'_{s_x^{(u)}} + i S'_{s_y^{(u)}}$ contains Pauli operators on all qubits except for $u$, where it contains a $2 P^{-} = X + i Y$ operator.
Moreover, since $S'_{s_x^{(u)}} \ket{0}^{\otimes N} = S_{s_x^{(u)}} \ket{0}^{\otimes N}$, and similarly for $S'_{s_y^{(u)}}$,
\begin{equation}
    \label{eq:vacuum_all_zeros}
    \left(S_{s_x^{(u)}} + i S_{s_y^{(u)}}\right) \ket{0}^{\otimes N} = 0.
\end{equation}
A similar argument for the creation operator can be used to show that $(S_{s_x^{(u)}} - i S_{s_y^{(u)}}) \ket{0}^{\otimes N}$ is proportional to a computational basis state: since $(S_{s_x^{(u)}} - i S_{s_y^{(u)}}) \ket{0}^{\otimes N} = (S'_{s_x^{(u)}} - i S'_{s_y^{(u)}}) \ket{0}^{\otimes N}$, its action is to flip qubit $u$'s state from $\ket{0}$ to $\ket{1}$, as well as possibly to flip any other qubits above $u$ in the tree for which the traversed links are labelled with an $X$ or a $Y$.
In Appendix \ref{app:prod_pres_proof}, we extend this argument to prove that any Fock basis state is mapped into a computational basis state in qubit space.

After application of this pairing scheme, the $j$-th mode operators take the form,
\begin{equation}
    \label{eq:general_form_tree_op}
    \begin{aligned}
            &a_j \mapsto \frac{1}{2} \left( X_{u} \prod_{k \in \mathcal{Z}^{x}_u} Z_k + i Y_u \prod_{k \in \mathcal{Z}^{y}_u} Z_k \right) G_u,\\
            &a_j^{\dagger} \mapsto \frac{1}{2} \left( X_{u} \prod_{k \in \mathcal{Z}^{x}_u} Z_k - i Y_u \prod_{k \in \mathcal{Z}^{y}_u} Z_k \right) G_u,
    \end{aligned}
\end{equation}
where $\mathcal{Z}^x_u$ and $\mathcal{Z}^{y}_u$ are sets of qubits that $S_{s_x^{(u)}}$ and $S_{s_y^{(u)}}$ act non-trivially on below qubit $u$ in the tree, and $G_u$ is a common Pauli string we can factor out. The sets $\mathcal{Z}^u_{x/y}$ may be empty, in which case the operator reduces to a similar form to the Jordan-Wigner mapping, $a_i^{(\dagger)} \rightarrow P_u^\pm G_u$, where $G_u$ enforces fermionic anti-symmetry with other qubits analogous to the $Z$-chain. This equation is graphically understood as $G_u$ being the common path of $S_{x/y}^{(u)}$ from the root to qubit-$u$, and sets $\mathcal{Z}^{x/y}_u$ are qubits along the Z-paths bifurcating from the X/Y-legs of qubit-$u$.

\begin{figure}[t]
    \centering
    \includegraphics[width=.95\columnwidth]{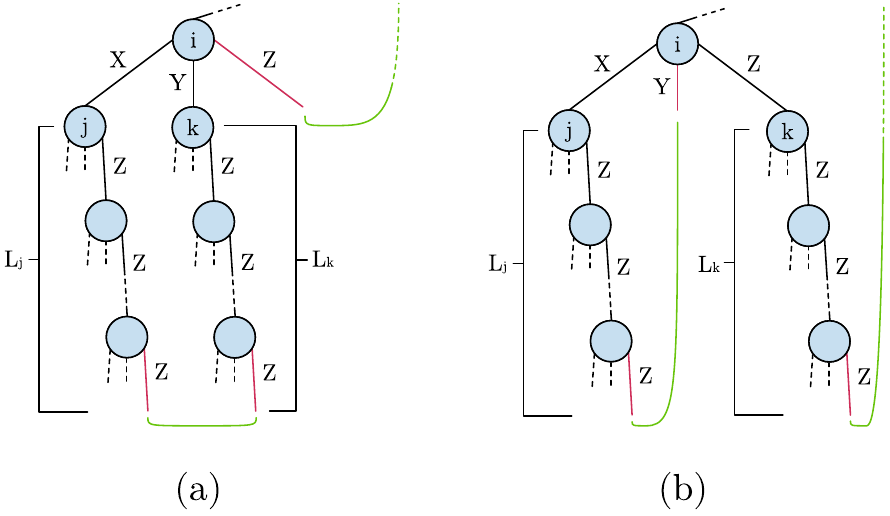}
    \caption{Figures depicting the effect of edge labelling below a node $i$.
    Its two descendants, $j$ and $k$, both have descendants, and the all-$Z$ paths underneath them have length $L_j$ and $L_k$, respectively.
    We assume that all links in the tree except for the ones right below $i$ have the same label in both figures (a) and (b).
    In the left figure, both paths contribute to occupation delocalisation of the mode associated with $i$, while in the right figure, only the path below $j$ does.
    In the latter case, the path of length $L_k$ may contribute to the delocalisation of some other mode unless the path from $i$ to the root only crosses $Z$-labelled edges.}
    \label{fig:labelling_choice}
\end{figure}

\subsection{Properties of the mappings and the effect of labelling}
\label{sec:mapping_properties}
In the construction of a TT mapping, there are two main degrees of freedom: the tree and the labelling.
In this subsection, we briefly discuss how the properties of these elements impact the resulting mappings.

An important feature of a fermion-to-qubit mapping is its \textit{Pauli weight}.
The Pauli weight of a Pauli string is defined as the number of qubits on which the string acts non-trivially (that is, with a Pauli operator different from identity).
In the case of a mapping, this refers to the Pauli weight of the strings of the mapped fermionic operators.
In the case of TT mappings, this can be easily analysed by considering the Majorana strings.
More concretely, notice that the Pauli weight of a Majorana string generated from a TT is precisely the length of the corresponding path from root to leg.
Thus, there is a straightforward connection between the Pauli weight of the mapping and the average shortest path length to the root in the tree.
More generally, the topology of the tree impacts the sets of qubits involved when applying creation or annihilation operators, but not how (i.e., by which Pauli operators).

The labelling, instead, has a more subtle impact on the resulting map.
While it cannot affect the average Pauli weight of the resulting Majorana strings, it impacts how the occupation of the fermionic modes is \textit{delocalised} over qubits.
More precisely, consider the number operator $n_j = a^{\dagger}_j a_j$ for a fermionic mode $j$ in qubit space.
Using Eq.~\eqref{eq:mapped_creation_annihilation}, we see that
\begin{equation}
    \label{eq:number_op_Majorana}
    n_j = \frac{1}{2} \left( \mathbbm{1} + i m_{2j} m_{2j + 1} \right) \leftrightarrow \frac{1}{2} \left( \mathbbm{1} + i S_{s_x^{(u)}} S_{s_y^{(u)}} \right).
\end{equation}
Since the two Pauli strings $S_{s_x^{(u)}}$ and $S_{s_y^{(u)}}$ are equal on all qubits above $u$ in the tree, the product on the right-hand side results in the identity operator for those.
On $u$, on the other hand, the product yields $XY = iZ$.
For the qubits below $u$ in the strings, however, one string acts with a $Z$ operator while the other one with identity.
Therefore, if we define the set $\mathcal{Z}_u$ of qubits below $u$ on which $S_{s_x^{(u)}}$ and $S_{s_y^{(u)}}$ act non-trivially, and include $u$ itself too, we have
\begin{equation}
    \label{eq:number_op_z}
    n_j = a^{\dagger}_j a_j \leftrightarrow \frac{1}{2} \left( \mathbbm{1} - \prod_{q \in \mathcal{Z}_u} Z_q\right),
\end{equation}
that is, the occupation of a mode $j$ is encoded in the parity of the state of the qubits in $\mathcal{Z}_u$.
Now, given a qubit $u$ with edges directly below itself, the choice of label for each of these edges will generally affect the structure of sets $\{ \mathcal{Z}_u \}_u$ in the resulting mapping and, with it, its delocalisation structure.

To analyse this in an illustrative manner, let us introduce a convenient definition of mode-specific delocalisation $D_u$ in terms of the qubit $u$ the mode is associated with in a mapping,
\begin{equation}
    \label{eq:def_delocalisation}
    D_u = | \mathcal{Z}_u | - 1.
\end{equation}
Now, consider a generic example in which a node $i$, which is not the root, has two descendants $j$ and $k$ (see Fig.~\ref{fig:labelling_choice}).
Nodes $j$ and $k$ themselves may have descendants.
Suppose that we add legs to all qubits and we label all links in the resulting tree except for the three links stemming downwards from $i$.
The question is then how those three links should be labelled.
Of course, there are three possibilities (three labels to be distributed among three links), but since swapping $X \leftrightarrow Y$ labels between two links stemming from the same node has a simple impact on the mapping (switching the roles of $S_{s_x^{(u)}}$ and $S_{s_y^{(u)}}$), the only two situations to be discussed are whether the $Z$ label should be assigned to an edge or the leg.

In the left figure, we depict the case in which the leg is assigned the $Z$ label, and the two edges, $X$ and $Y$.
In this case, $\mathcal{Z}_i$ contains $i$ and all the nodes in the $Z$-strings lying below $j$ and $k$, which in the example have length $L_j$ and $L_k$, respectively.
Therefore, the occupation of the mode associated with qubit $i$ is delocalised among $D_i = L_j + L_k$ qubits.
In the opposite case, in which the $Z$ label is assigned to one of the edges, on the other hand, one of the two $Z$-strings no longer contributes to the delocalisation of node $i$.
In the illustration, we have $D_i =L_j$, that is, the occupation of the mode is less spread in the second case.

\begin{figure*}[t]
   \centering
   \includegraphics[width=0.8\linewidth]{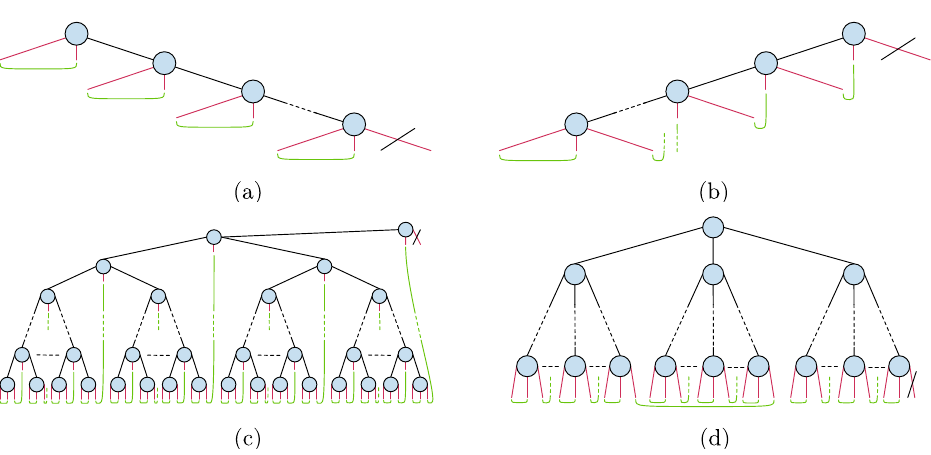}
   \caption{Ternary trees generating four paradigmatic mappings.
   a) Jordan-Wigner, b) Parity, c) Bravyi-Kitaev and d) JKMN mapping.
   From the analysis presented throughout Sect.~\ref{sec:mapping_properties}, we can see that both Jordan-Wigner and Parity have linear-scaling depth, and hence Pauli weight, while Bravyi-Kitaev and JKMN have $O(\log_2 N)$ and $O(\log_3 N)$ scaling, respectively.
   Regarding the delocalisation, it can be seen that Jordan-Wigner minimises it, while Bravyi-Kitaev and Parity have no nodes other than the root in the all-$Z$ branch.
   JKMN has a non-minimal all-$Z$ branch including $O(\log_3 N)$ nodes, and hence is nearly maximally delocalised for large $N$.}
   \label{fig:paradigmatic_mappings}
\end{figure*}

However, notice one important fact: assume we follow the path from $i$ upwards towards the root $r$ until we reach an edge labelled with an $X$ or a $Y$, and let us call $u$ the node reached by traversing that edge.
In the latter case (right-hand side of the figure), the $Z$-string along qubit $k$ is now part of one of the $Z$ strings directly below qubit $u$.
In other words, while the delocalisation of the mode associated with $i$ has decreased by an amount $L_k$, the delocalisation of the mode associated with $u$ has increased by the same amount.
Therefore, under this assumption, the labelling cannot affect the average delocalisation of the mapping, but only its distribution among the qubits.
Crucially, if the path from $i$ to $r$ only crosses $Z$-labelled edges, this is no longer true, and the second labelling does not increase the delocalisation of any other mode.

From the above discussion, we can draw a very useful overall conclusion regarding the delocalisation structure of a mapping: the average delocalisation among nodes is given by
\begin{equation}
    \label{eq:avg_delocalisation}
    \left\langle D_u \right\rangle = 1 - \frac{h_Z}{N},
\end{equation}
where $h_Z$ is simply the number of nodes that can reach the root node $r$ by traversing only $Z$-labelled edges, including the root itself.
This can be seen as follows.
If a node $i$ cannot be traced back to the root following $Z$-labelled edges, the path towards $r$ must cross an $X$- or $Y$-labelled edge attached to some node $u$, and thus $i$ contributes one unit to the delocalisation of node $u$.
Therefore, the sum of all delocalisations must be equal to the number of nodes not in the $Z$-labelled path, that is, $\sum_u D_u = N - h_Z$.
This observation implies that, in order to minimise the delocalisation of the modes, which may be a desirable property of a fermion-to-qubit mapping \cite{sawaya2016error}, we must maximise the number of nodes along the $Z$-only path.
Interestingly, since $h_Z \in \{ 1, \ldots, N \}$, the average delocalisation is bounded $\left\langle D_u \right\rangle \in [0, 1 - 1/N]$, that is, on average, the occupation of the fermionic modes is stored in less than two qubits, $\left\langle | \mathcal{Z}_u | \right\rangle \in [1, 2 - 1/N]$.

\subsection{The ternary trees of paradigmatic mappings}
It is illustrative to analyse paradigmatic fermion-to-qubit mappings in this context.
In particular, Jordan-Wigner (JW), Bravyi-Kitaev (BK), Parity (P), and obviously the optimal mapping from Ref.~\cite{jiang2020optimal} (JKMN) are all 1-NTO and can be generated from TT.
In Fig.~\ref{fig:paradigmatic_mappings} we depict their corresponding trees.
By analysing their graph topologies, and following the insights from the previous discussion, we can easily understand their main properties.

Both JW and P are given by linear graphs.
Since these are depth-$N$ trees, the Pauli weight of the resulting Majorana strings is $O(N)$.
However, their occupation delocalisation is different.
JW is an extreme case, given that all nodes are in the $Z$-labelled path and thus has average delocalisation $\left\langle D_u \right\rangle = 0$; this is the only possible TT mapping with no delocalisation.
In the case of P, the occupation is maximally delocalised, with each occupation encoded between two consecutive nodes in the chain, except for the last qubit, in which it is fully localised.

\begin{figure*}[t]
    \centering
    \includegraphics[width=0.7\textwidth]{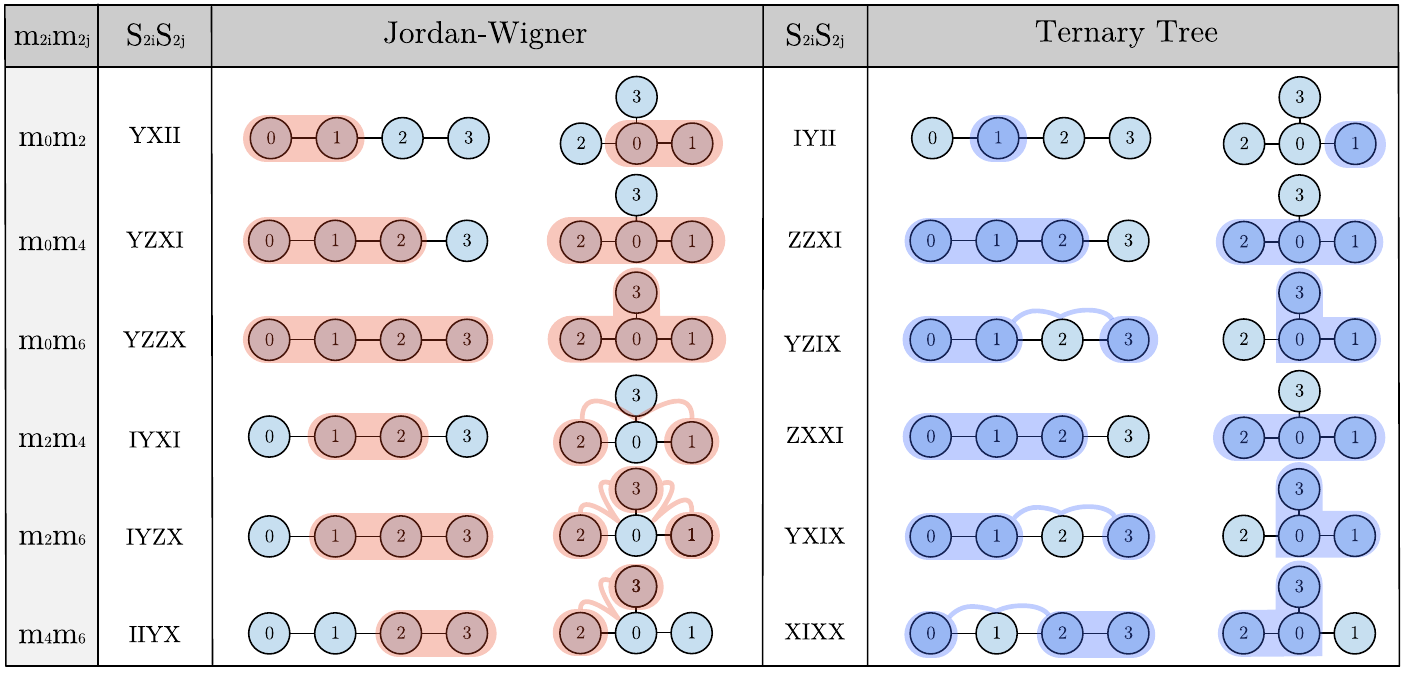}
    \caption{Qubit operators resulting from the product of two even Majorana operators, $m_{2i} m_{2j}$, implemented on a linear and a clustered topology.
    The operators are mapped to qubit space using JW and a four-qubit TT mapping in which the root has degree three.
    Each product of Majorana operators yields a Pauli string, which is written explicitly in the leftmost column of each mapping.
    The red (blue) highlight corresponds to qubits on which the Pauli strings act non-trivially.
    A highlighted line skipping over qubits denotes ones that are not present in the strings but are involved in the cascade of SWAP gates.
    }
    \label{fig:swap_overhead_example}
\end{figure*}

BK and JKMN, instead, are generated by trees with constant branching rates 2 and 3, respectively.
Therefore, their depths, and hence the resulting Pauli weights, scale as $O(\log N)$, with JKMN having a smaller depth owing to its higher branching rate (in fact, the authors prove the optimality of the Pauli weight of their mapping in Ref.~\cite{jiang2020optimal}).
In both cases, however, the price to pay is the delocalisation of the occupation.
More precisely, notice that the higher up the tree the common ancestor of a given pairing is, the more $Z$-links are involved in the resulting Majorana strings.
Thus, only the lowest-lying nodes lead to completely localised modes.

\section{Growing hardware-efficient mappings with the bonsai algorithm}
\label{sec:hardware_efficient_mappings}
Ternary tree mappings can be used as a tool for the design of custom fermion to qubit mappings.
The framework introduced in Sect.~\ref{sec:mappings_from_trees} is general and can help find mappings with specific desired properties by tailoring the trees according to different cost functions.
In what follows, we introduce an algorithm to produce mappings aimed at reducing the complexity of fermionic simulations on quantum computers by minimising the impact of limited qubit connectivity in the quantum processor.
We start this section by briefly introducing the problem, and we then present the Bonsai algorithm along with an illustrative and important use case: heavy-hexagon qubit lattices, the topology of choice for current IBM quantum computers.

\subsection{Fermionic simulation under limited connectivity constraints}
\label{sec:limited_connectivity}
The simulation of fermionic many-body systems is one of the most promising applications of quantum computing, both in the near term and in the fault-tolerant era.
Many of the existing algorithms work in second quantisation, and thus typically require mapping the fermionic operators into qubit space.
Fermionic operations are then mapped to unitary gates among the qubits in the device.
However, many platforms (such as superconducting qubits) have limited connectivity, meaning that many pairs of qubits in the processor cannot physically interact directly.
Thus, when a quantum gate involves qubits that are not physically connected, SWAP gates are iteratively applied so that the state of distant qubits are transported to neighbouring ones, and the gate is then applied.
While this is always possible in theory, in practice, the additional SWAP gates increase the circuit complexity, which results in longer runtimes and, consequently, the increased detrimental effect of noise.

In order to illustrate how limited connectivity impacts the circuit complexity, let us consider a minimal example with four fermionic modes simulated with four qubits on two different platforms, one with linear connectivity (the physical connectivity graph being a one-dimensional chain), and a second one with star-like connectivity (three of the four qubits connected to the fourth, and no other connections).
Both are depicted in Fig.~\ref{fig:swap_overhead_example}.
We now map four fermionic modes to these qubits using JW and TT mapping.
For the latter, we consider the specific situation in which the ternary tree is congruent with the connectivity graph of the qubits: qubit 0 is the root, and the other three qubits are its descendants.

\begin{figure*}[t]
    \centering
    \includegraphics[width=0.95\textwidth]{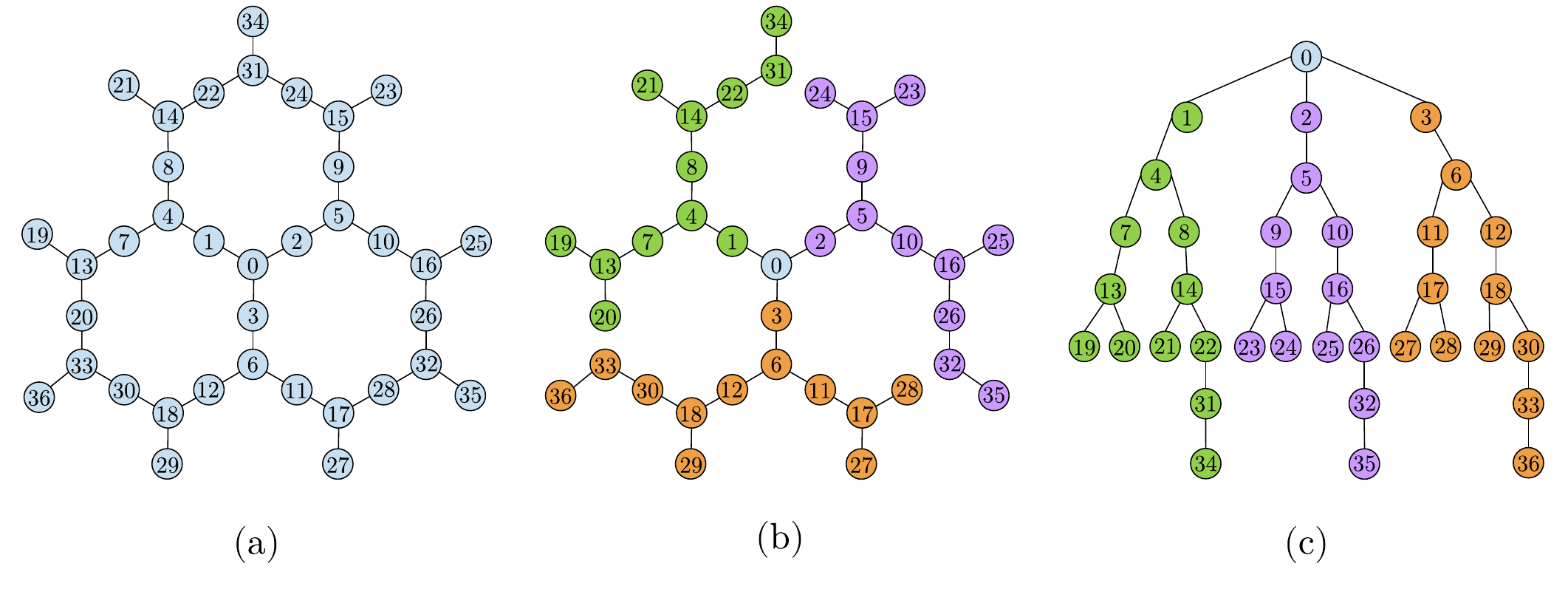}
    \caption{a) Connectivity graph of a 37-qubit heavy-hexagon processor.
    b) Unfolded tree, illuminating how the paths of the tree follow the hardware topology in order to create a hardware-efficient mapping.
    The colouring of vertices indicates where the three initial branches from the root qubit unfold to.
    c) The same tree as in b), folded in the layout used in the rest of this work.}
    \label{fig:heavy_hexagon_st}
\end{figure*}

We now examine the simulation of the even-even Majorana terms, $m_{2i} m_{2j}$ for $i\neq j$, arising from single excitation terms, $a_i^{\dagger} a_j$.
The resulting six Pauli strings are tabulated in the figure.
In many applications, these terms must be exponentiated and implemented as rotations, i.e., $\exp{-i\theta m_{2i} m_{2j}}$.
This requires entangling gates between the qubits not acted upon by an identity in the corresponding Pauli string.
As explained above, if two such qubits are not neighbours in the physical connectivity graph of the device, SWAPs must be applied.
Figure \ref{fig:swap_overhead_example} highlights the qubits involved in implementing the rotations.
The SWAP overhead is indicated by thin lines skipping over the qubits.

In Fig.~\ref{fig:swap_overhead_example}, we observe that with TT and linear connectivity, three operators, $m_{0}m_{6}, m_{2}m_{6}, m_{4}m_{6}$, involve all four qubits even though the actual Pauli strings only act on three qubits each.
The JW mapping, on the other hand, is more congruent with the underlying connectivity.
In the large $N$ limit, the regular TT mapping presents an advantage in terms of Pauli weight with respect to JW (the former scales as $O(\log_3 N)$, while the latter as $O(N)$) so, in principle, each such rotation would involve much fewer qubits.
However, the SWAP overhead with limited connectivity reduces the Pauli weight advantage for the regular TT (and similarly for BK) and an amount of CNOTs equivalent to JW may typically be required.
In the case of star-graph connectivity, on the other hand, the TT mapping never requires SWAPs, as opposed to JW, and moreover, no operation involves more than three qubits.

This simple example illustrates why limited connectivity can be an issue for the implementation of fermionic operations, and also that the right choice of mapping, in particular one that is congruent with the underlying connectivity, can help mitigate the overhead.

\subsection{The Bonsai algorithm}
\label{sec:bonsai_algorithm}
In this section, we introduce an algorithm to generate custom fermion-to-qubit mappings tailored to device-specific connectivity graphs.
More precisely, the problem is, given a quantum processor, to find a mapping such that: 1) it is product-preserving, 2) the resulting Pauli weight is low, and 3) mode occupancy is local in qubit space.
The first condition is satisfied by appropriate pairing as described in Sect.~\ref{sec:pairing}.
The second and third points are suitably satisfied by finding a ternary tree that is a subgraph of the physical connectivity graph (or close to one) and then exploiting the labelling freedom to define how the mode occupancy is distributed over qubits in a rational manner.
In the following, we present this heuristic strategy in detail and illustrate it with an important application: designing mappings for heavy-hexagon quantum computers.
The steps of the algorithm are summarised in Algorithm \ref{alg:bonsai}, while the specific subroutines are described in detail in Appendix \ref{app:algorithms}.

\textit{Finding the ternary tree.}
The input of the Bonsai algorithm is a physical connectivity graph $\mathcal{P} = (V_\mathcal{P}, E_\mathcal{P})$, in which the nodes are the qubits in the processor and the edges represent the pairs of qubits onto which it is possible to physically apply entangling gates.
In Fig.~\ref{fig:heavy_hexagon_st} a), we depict the physical connectivity graph $\mathcal{P}$ of a 37-qubit heavy-hexagon computer.
Now, the strategy to minimise the SWAP overhead is to find a TT, $\mathcal{T} = (V_\mathcal{T}, E_\mathcal{T})$, that is congruent with the topology of $\mathcal{P}$.
More precisely, suppose that $\mathcal{T}$ is a subgraph of $\mathcal{P}$ (that is, $V_\mathcal{T} = V_\mathcal{P}$ and $E_\mathcal{T} \subseteq E_\mathcal{P}$).
Then, any path from the root to leaf in $\mathcal{T}$ is a path in $\mathcal{P}$ and, consequently, no SWAPs are required to apply a gate generated by a Majorana string.
A similar argument can be used for gates generated by single-excitation operators.

\begin{figure*}[t]
    \centering
    \includegraphics[width=.95\linewidth]{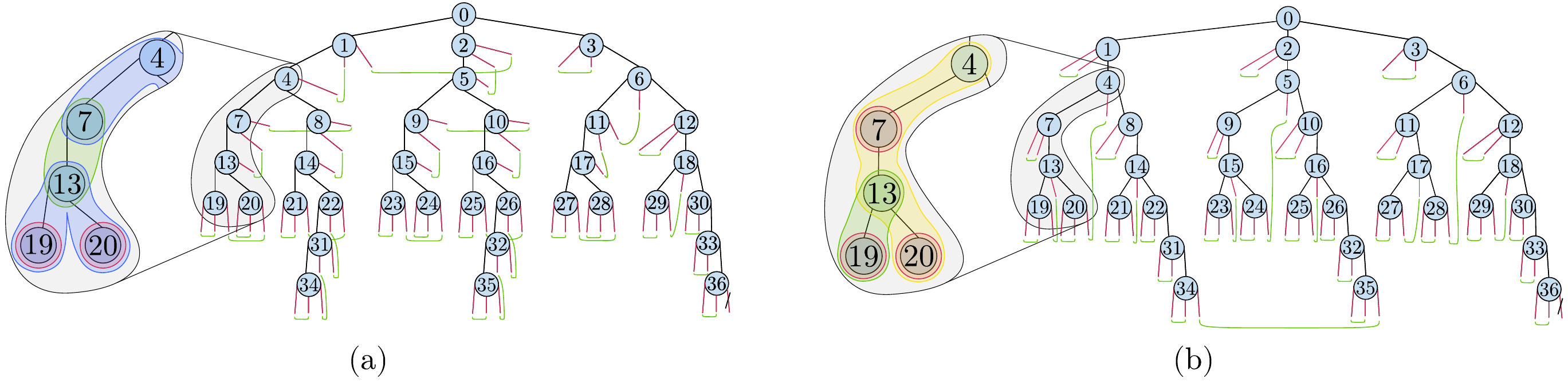}
    
    \caption{Mappings resulting from two different labelling strategies applied to the tree in Fig.~\ref{fig:heavy_hexagon_st}.
    a) By applying the homogeneous localisation labelling, modes associated with qubits with descendants are delocalised in a rather even fashion: nodes with one and two descendants have delocalisation $D_u = 1$ and $D_u = 2$, respectively.
    The zoomed-in area involving qubits 4, 7, 13, 19, and 20 further illustrates the occupancy distribution.
    The modes associated with 4 and 13 involve three qubits, the mode in qubit 7 involves 2, and the ones in 19 and 20, only one.
    b) The application of the heterogeneous localisation strategy yields a very different delocalisation structure.
    Nodes with one descendant are completely localised in this case ($D_u = 0$), but nodes with two descendants can be fairly delocalised.
    For instance, looking again at the four nodes in the shaded area, we see that the modes in qubits 7 and 13 are more localised than in the previous case, while the one in qubit 4 is delocalised among more qubits, with $D_4 = 3$.
    The particular case of the root node is a clear example, as it now exhibits delocalisation $D_0 = 14$.}
    \label{fig:mode_op_tree_folded}
\end{figure*}

A tree subgraph $\mathcal{T}$ that spans all the nodes in a graph $\mathcal{P}$ is called a \textit{spanning tree} (ST).
If $\mathcal{P}$ is a tree itself, then the choice of ST is unique.
A general graph, however, may have several STs.
A degree-$\Delta$ constrained ST is one such tree that has no vertices with a degree greater than $\Delta$.
In our case, since we need the subgraph $\mathcal{T}$ to be a ternary tree in order to define a mapping, all nodes but one must be at most degree-4 and the root degree-3.
This implies that it is not always possible to find such a tree (for instance, if $\mathcal{P}$ is a tree but not degree-$\Delta$ constrained with $\Delta \leq 4$).
Moreover, even if a degree-4 constrained tree subgraph exists, finding it is generally hard (in fact, simply determining whether there is one is an NP-complete problem \cite{garey1979computers}).
For our purposes, if $\mathcal{T}$ is not an ST of $\mathcal{P}$, it can nevertheless define a proper fermion-to-qubit mapping, although in such case some SWAPs may be needed to implement Majorana-generated unitary gates.
Therefore, we propose using a greedy heuristic to find a TT that is close to a spanning tree and is in fact guaranteed to find an ST for some specific topologies.
The routine is explained precisely in Appendix \ref{app:algorithms} (Algorithm \ref{alg:spanning_tree}).

\LinesNotNumbered
\begin{algorithm}[t]
    \label{alg:bonsai}
    \caption{Bonsai algorithm}
    
    Find a ternary tree $\mathcal{T}$ congruent with the physical connectivity graph $\mathcal{P}$ using Algorithm \ref{alg:spanning_tree}, which consists of the routines:
    
        \begin{itemize}
        
            \item[a.] Find a degree-constrained tree subgraph $\mathcal{T}$ using greedy search throughout $\mathcal{P}$.
            
            \item[b.] If the resulting tree does not span all nodes, add the remaining ones connecting them as to minimise the physical distance to nodes already in $\mathcal{T}$.
            
        \end{itemize}
        
    Add legs and introduce labels to $\mathcal{T}$ using Algorithm \ref{alg:branching_sub}, choosing among
    
        \begin{itemize}
        
            \item[a.] \textit{Homogeneous localisation}: occupancy is spread evenly over qubits in the tree.
            
            \item[b.] \textit{Heterogeneous localisation}: a subset of mode operators will act on many qubits while reducing the amount that others act on.
            
        \end{itemize}
        
    Pair the generated strings using Algorithm \ref{alg:pairing_scheme}.
    
\end{algorithm}

The idea is to start by defining $\mathcal{T} = (V_\mathcal{T}, E_\mathcal{T})$, with empty $V_\mathcal{T}$ and $E_\mathcal{T}$, and grow the tree iteratively.
First, choose a node to be the root $r$ of the TT, and define $\mathcal{L}_0 = {r}$.
The choice of the root has an impact on the resulting Pauli weight and average delocalisation of the mapping, as will be discussed later on; we now choose it to be central in $\mathcal{P}$ (that is, such that it minimises the distance to its furthest node, $r = \text{argmin}_u \max_v d(u, v)$, where $d(u, v)$ is the topological distance between nodes $u$ and $v$ in $\mathcal{P}$).
In Fig.~\ref{fig:heavy_hexagon_st} b), this is the pale blue central node.
Next, define an empty set $\mathcal{L}_1$, and add to it $\min(\Delta(r), 3)$ neighbours of $r$ in $\mathcal{P}$.
For every node $u$ that is added, add the link between $r$ and $u$ to $E_\mathcal{T}$.
Notice that $r$ may have degree $\Delta(r) > 3$.
In that case, the choice is not unique.
For simplicity, we suggest choosing three of them randomly.
Then, the process is repeated for each node in $\mathcal{L}_1$: define $\mathcal{L}_2 = \emptyset$ and add to it up to three neighbours of each node in $\mathcal{L}_1$ that have not yet been added, that is, not in $\mathcal{L}_0 \cup \mathcal{L}_1 \cup \mathcal{L}_2$, and the corresponding links to $E_\mathcal{T}$.
By iterating this process, at some point, all neighbours of all nodes in $\mathcal{L}_L$ for some $L$ have been added to some $\mathcal{L}_i$, so the procedure must stop.
Now, let $V_\mathcal{T} = \bigcup_{i=1, \ldots, L} \mathcal{L}_i$.
If $V_\mathcal{T} = V_\mathcal{P}$, we have found a degree-4 constrained ST of $\mathcal{P}$.
Notice that this procedure succeeds with heavy-hexagon lattices, as shown in Fig.~\ref{fig:heavy_hexagon_st} b) and c).

If the above procedure does not span all the qubits in $\mathcal{P}$, we need to add the remaining nodes in $V_\mathcal{P} \setminus V_\mathcal{T}$ to $\mathcal{T}$ according to some criterion.
Notice that it is always possible to include these nodes in $\mathcal{T}$ through ``virtual edges'' that connect physically detached nodes at the expense of SWAPs in the compilation.
In order to minimise the resulting SWAP overhead, a good strategy is trying to minimise the physical topological distance between qubits connected in $\mathcal{T}$.
This can be achieved following a greedy criterion: for every node $u$ in $V_\mathcal{P} \setminus V_\mathcal{T}$, find amongst the nodes $v$ in $V_\mathcal{T}$ with a downward degree less than three in $\mathcal{T}$ the ones that minimise the distance $d(u, v)$ in $\mathcal{P}$, and connect $u$ to one of them.

It is worth noting a few aspects of this method.
On the one hand, since at each step in the first part of the algorithm we add as many neighbours of each node as the topology allows, we are implicitly minimising the depth of the resulting tree.
Indeed, notice that if the physical device is all-to-all connected, then the resulting graph is the TT from Ref.~\cite{jiang2020optimal} with optimal depth $\mathcal{O}(\log _3 N)$.
In the case of the heavy-hexagon topology, the greedy algorithm succeeds in finding degree-constrained spanning trees with depth scaling as $O(\sqrt{N})$, i.e., with quadratically lower Pauli weight than Jordan-Wigner.
This latter point can be seen through geometric arguments: the number of qubits at a given topological distance smaller than $R$ from a chosen root node $r$ scales as $R^2$.
On the other hand, if the connectivity graph is a chain, the algorithm, as presented above, would choose as root $r$ a node in the centre of the chain.
While this would lead to a mapping with Pauli weight lower than JW, the occupation would be more delocalised than in the latter case.
Instead, if one is interested in minimising delocalisation, a better choice of the root is a node that lies on an extreme of a \textit{diameter} of $\mathcal{P}$ (that is, one of its longest shortest paths), so that the edges along the longest shortest path can be later labelled with $Z$, hence maximising $h_Z$.
In such a case, JW would be obtained for a chain.
In general, this trade-off between Pauli weight and delocalisation can be easily controlled with the choice of the root node.

\textit{Labelling the tree.}
Once the TT has been identified, the next step is to introduce terminating legs and Pauli labels to the links to create a qubit tree.
As discussed in the previous sections, in order to minimise the average delocalisation, we must label with $Z$ all edges from the root to the most distal node from it.
In the case of the heavy-hexagon in Fig.~\ref{fig:heavy_hexagon_st}, we may do so with all the edges between nodes 0 (the root) and 36.
Using the different labelling techniques, we can decide to a certain degree how mode occupancy localisation is spread.
While these cannot affect the average delocalisation, $\langle D_u \rangle$, they can determine how heterogeneously distributed among the qubits the occupancy can be.
To that end, we introduce two different labelling strategies, which we coin \textit{homogeneous} and \textit{heterogeneous} localisation, based on the discussion in Sect.~\ref{sec:mapping_properties}.

\begin{figure*}[t]
    \centering
    \includegraphics[width=0.85\linewidth]{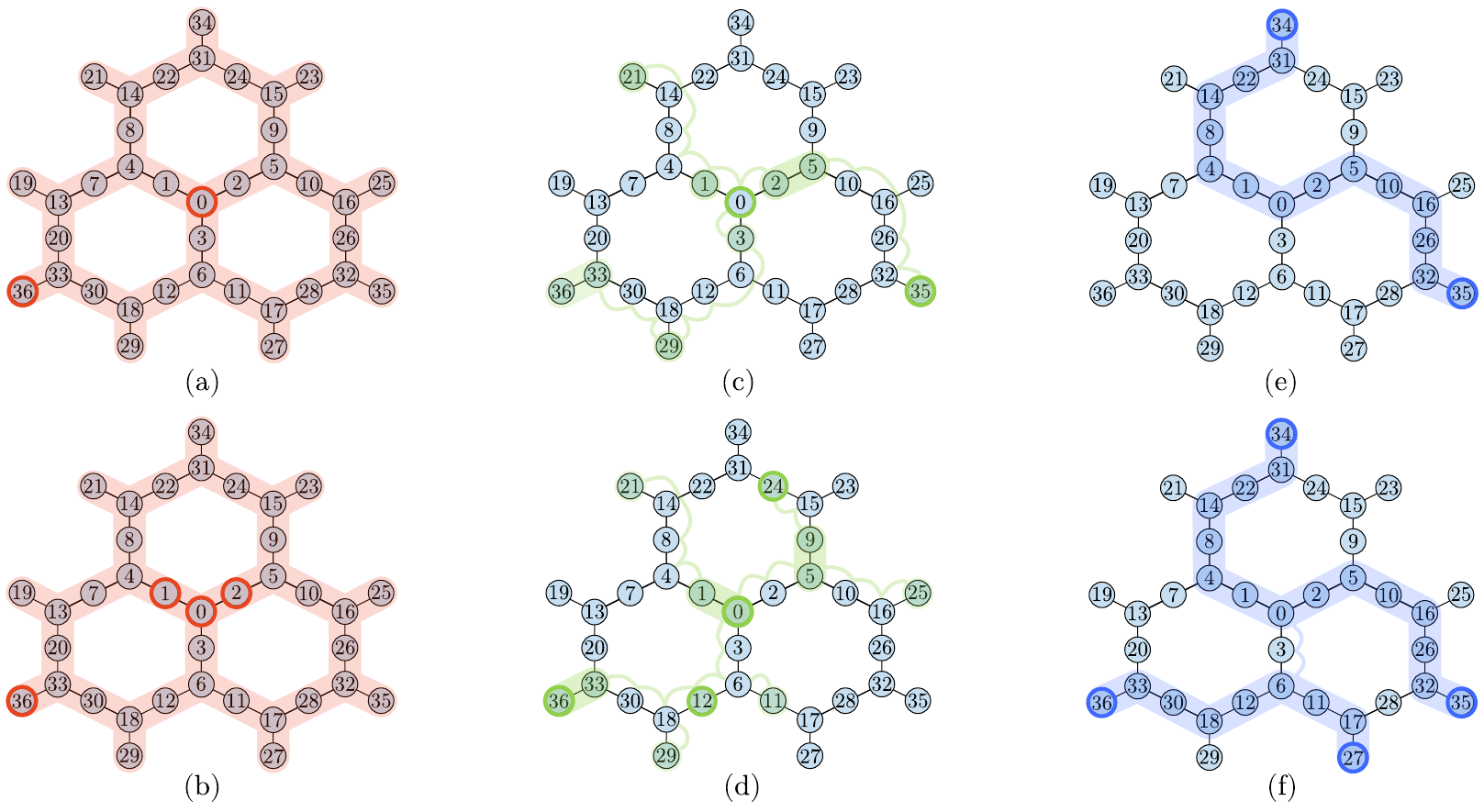}
    \caption{The highlight indicates qubits involved in application of worst-case excitations, both for one- and two-particle terms.
    The top (a, c, e), and bottom (b, d, f) rows correspond to single and double excitation terms respectively.
    For simplicity, we assume that the identification between modes $j$ and qubits $u$ in Eq.~\eqref{eq:pairing_identification} is such that $j = u$. Qubits to which modes are associated are circled in red, blue, and green for the JW, BK, and the custom map. For example, (a) and (b) correspond to qubit operators of modes-$(0,36)$ and modes-$(0,1,2,36)$ accordingly.
    In the case of JW (a, b), both excitations result in gates acting on all qubits. For BK (c, d) the resulting gates will act on many disconnected qubits, resulting in a high SWAP cost completely mitigating its logarithmic scaling benefit. 
    The custom mapping (e, f) generated by the Bonsai algorithm, on the other hand, presents much simpler worst-case scenarios.
    The single- and double-excitation terms involve much fewer qubits than JW, and only two SWAPs are required to connect the separate highlighted regions in the latter case.}
    \label{fig:heavy_hexagon_worst_cases}
\end{figure*}

Homogeneous localisation proceeds by maximising the number of $XY$ branches amongst edges stemming from the same nodes, in a similar fashion as in Fig.~\ref{fig:labelling_choice} (left).
Thus, pairs of edges below a node are assigned $X$ and $Y$ labels, while single edges have an $X$ label.
Heterogeneous localisation, instead, maximises the number of $Z$ labels amongst edges: if a node has two descendants, one of the edges is assigned a $Z$ and the other one an $X$.
If the node has only one edge, it is assigned a $Z$.
In this way, the heterogeneous localisation assignment tends to localise the occupation of single-edge nodes at the expense of the delocalisation of nodes above them, hence resulting in typically more heterogeneous distributions of localisation.

In Fig.~\ref{fig:mode_op_tree_folded}, we depict the two labelling outcomes for the heavy-hexagon topology.
Homogeneous localisation produces many operators with occupancy depending on at worst case three qubits, giving a typical delocalisation $D_u = 2$ for those.
This is reflected also in the fact that the green lines representing the pairings do not span distant qubits.
The number of completely local operators, with specific delocalisation $D_u = 0$ is 16 in this case, whereas in the case of heterogeneous localisation, 27 operators are fully localised.
This has the side effect of creating fewer operators with higher delocalisation, such as the number operator for the mode associated with the root (qubit 0), with delocalisation $D_0 = 14$. 
Table \ref{tab:hh_mode_ops} in appendix ~\ref{app:HH mode ops} exhibits mode operators generated for both localisation schemes in this heavy-hexagon example.

Choosing one delocalisation scheme or the other is application-specific and will alter the structure of the resulting circuit. 
This is evident when considering a highly delocalised mode operator. 
Operations derived from this mode will generally act on more qubits than its localised counterpart, consequently resulting in more expensive circuits. 
However, since every tree-mapping has a certain level of average delocalisation, careful handling is required. 
In certain cases, a few modes may hold little importance and can be delocalised without significant detrimental consequences. 
In such scenarios, employing the heterogeneous localisation strategy to localise the relevant modes can prove beneficial.
This will lead to an overall improvement as frequently used modes become less delocalised. 
Conversely, in cases where assumptions about the structure cannot be made, applying homogeneous localisation may prove a safe option. 
This strategy uniformly spreads the delocalisation, ensuring a balanced distribution across the modes.

With these mappings, circuit cost is reduced in two ways.
Given that Majorana products follow paths along the paths in the tree structure, the number of SWAP gates needed for single excitations is zero, and the number is diminished for double excitations due to Majorana products following paths along the tree structure.
The number of entangling gates is further mitigated by lessening Pauli weight.
This is illustrated in Fig.~\ref{fig:heavy_hexagon_worst_cases}, where we highlight the interaction maps of worst-case single and double excitations for the mapping obtained through homogeneous localisation, compared to the JW and BK mappings.
In both cases, JW acts extensively on the whole system. 
For BK the qubits involved are disconnected and will need many SWAPs to compile, mitigating the circuit benefits of the encoding's logarithmic Pauli-weight scaling.
This is not the case for our custom encoding, where the reduction is approximately two-thirds of the system for single and one-third for double excitations.
For the latter case, two SWAP gates are required to bridge across the disconnected interaction regions.

Another interesting aspect of these custom mappings is that it simplifies the transpilation of the circuits.
Since the mapping is designed to be congruent with the hardware, it is not necessary to search for the optimal qubit assignment, but only to solve the Steiner graph problem to determine where SWAP gates are required. In addition, for two-dimensional devices other than the heavy-hexagon based connectivity studied here (e.g., Google's Sycamore grid topology \cite{google_quant_supremacy}), we expect a similar square root scaling. 

It is noteworthy that the formalisms presented in this study can be expanded to tackle the difficulties posed by fully connected devices like ion-traps. This could involve exploring modified mappings to minimize the Hamiltonian Pauli weight linked to reduced measurement cost \cite{garcia2021learning}, or devising mappings based on the circuit's structure rather than the hardware to mitigate costs. However, such investigations are deferred to future research.

\section{Conclusions}
\label{sec:conclusions}
In this work, we have considered fermion-to-qubit mappings relying on the identification of sets of Pauli strings obeying the anti-commutation relations of Majorana operators.
Within this context, we have focused on a specific class, arguably the simplest one to work with, in which the Pauli strings have a non-trivial overlap involving just one qubit.
We have then presented a framework that enables sampling such mappings while designing many of their resulting properties.
An important element of the methodology is the pairing algorithm that ensures the preservation of separability, that is, that uncorrelated fermionic states are mapped to uncorrelated qubit states. 
Interestingly, the framework contains paradigmatic mappings as particular instances, which allows us to interpolate between them at will.

With this framework at hand, we have devised an algorithm to design hardware-specific mappings with lower SWAP overhead than other paradigmatic mappings while retaining a fair localisation of the fermionic occupation in qubit space.
When applied to the heavy-hexagon architecture, we obtain a mapping with a quadratically lower Pauli weight than JW.
Importantly, the mapping enables applying single excitation operations with no SWAP overhead, and double excitations with a minor one.
This can result in a significant improvement in circuit complexity with respect to hardware-agnostic mappings.

Currently, JW is the mapping of choice in most simulations on limited connectivity hardware \cite{bentellis2023benchmarking,qsimmat,motta2021low,doi:10.1126/science.abb9811}, partly due to the fact that its linear generating tree structure (see Fig.~\ref{fig:paradigmatic_mappings}) makes it easy to find a set of qubits with that connectivity within the device.
For other mappings like BK or TT, suitable subgraphs that have tree-like structure are unattainable on limited connectivity, resulting in a SWAP overhead negating the logarithmic advantage.
Bonsai encodings, on the other hand, enable leveraging higher dimensions of limited connectivity graphs to reduce simulation cost, as they naturally extend the hardware suitability of JW while reducing the non-locality of the mapping.

The versatility of the approach here presented enables many other possibilities.
In terms of designing mappings, the choice of the cost function to be optimised for is not unique, so the Bonsai algorithm can be naturally extended to produce encodings with different desirable properties.
In particular, an important application is extracting relevant physical quantities of the system using local informationally complete POVMs \cite{jiang2020optimal, garcia2021learning, glos2022adaptive}.
In this case, the Pauli weight of the observable is the dominant figure of merit, which is why the authors proposed the logarithmic-depth regular ternary tree in  Ref.~\cite{jiang2020optimal}.
While that is the optimal mapping in terms of measurement cost for arbitrary fermionic reduced density matrix elements, in practice, one is typically interested in specific observables like the energy.
In that case, the mapping may be further optimised to reduce the measurement cost of e.g.~the Hamiltonian of the system.
Moreover, it would be interesting to do so while limiting the incurred SWAP overhead on specific hardware.

In broader, more theoretical terms, we emphasise that the bulk of the work here presented is devoted to a specific subset of all the possible mappings, the 1-NTO class, which includes all the widely used encodings.
As we have proved, with the pairing we introduced, any root-containing connected ternary tree yields a valid, product-preserving fermion-to-qubit mapping.
However, we have also shown with a counter-example that not all 1-NTO maps can be generated in this fashion, so the question of how to characterise and represent the space of 1-NTO encodings remains open.
In addition, as noted in Section \ref{sec:theory}, $k$-NTO maps with $k > 1$ do exist.
This opens the interesting prospect of studying these somewhat exotic mappings.

\noindent {\bf Additional information.} The Bonsai algorithm is part of \emph{Aurora}'s suite of algorithms for chemistry simulation.

\appendix

\section{Examples of exotic fermion-to-qubit mappings}\label{app:exotic}

The most used fermion-to-qubit mappings, such as the JW, BK, and Parity mappings are all 1-NTO mappings, and can even be generated from ternary trees. In this Appendix, we provide a toy example of a fermion-to-qubit mapping that is not 1-NTO and another one that is 1-NTO but cannot be generated from ternary trees.

Consider the following mapping of a four-mode fermion system to a 4-qubit system:
\begin{align*}
    &m_0  \mapsto  X_1 X_2 X_3, \quad  &&m_1  \mapsto  Y_1 Y_2 Y_3, \\
    &m_2  \mapsto  X_0Z_1 Y_2 Y_3, \quad  &&m_3  \mapsto  Y_0Z_1 X_2 X_3, \\
    &m_4  \mapsto  Y_0 Y_1 X_3, \quad  &&m_5  \mapsto  X_0 X_1 Y_3, \\
    &m_6  \mapsto  X_0 X_1 X_2Z_3, \quad  &&m_7  \mapsto  Y_0 Y_1 Y_2Z_3. \\
\end{align*}
One can easily check that the mapping satisfies Criteria (A)-(C) of Subsection~\ref{subsec:MString}, thus it is a valid Majorana string mapping. The non-trivial overlap between the Majorana strings $X_1 X_2 X_3$ and is $3$, thus this cannot be a 1-NTO mapping, but is instead 3-NTO.

An example of a 1-NTO Majorana string mapping (for 3 fermionic modes), which cannot be generated from a ternary tree is the following:
\begin{align*}
    &m_0  \mapsto  X_0 Z_1, \quad  &&m_1  \mapsto  Y_0 Z_1, \\
    &m_2  \mapsto  X_1 Z_2 , \quad  &&m_3  \mapsto  Y_1 Z_2, \\
    &m_4  \mapsto  Z_0 X_2, \quad  &&m_5  \mapsto  Z_0 Y_2. \\
\end{align*}

\section{Algebraic independence of subsets of TT-generated Pauli strings}\label{app:alg_ind_proof}
The aim of this section is to prove that any subset $\mathcal{S}' \subset \mathcal{S}$ ($\vert \mathcal{S}' \vert < \vert \mathcal{S} \vert$) of the set $\mathcal{S}$ of $2N + 1$ Pauli strings generated by an $N$-node TT is algebraically independent, that is, that there are no two different subsets $A \subseteq \mathcal{S}'$ and $B \subseteq \mathcal{S}'$, $A \neq B$, such that $\prod_{S_i \in A} S_i \propto \prod_{S_j \in B} S_j$.

First, notice that it is enough to prove that no two \textit{disjoint} subsets $A$ and $B$ leading to equal products exist, given that
\begin{equation}
    \label{eq:irrelevant_intersection1}
    \prod_{S_i \in A} S_i \propto \prod_{S_j \in B} S_j \Rightarrow \prod_{S_i \in A \setminus A \cap B} S_i \propto \prod_{S_j \in B \setminus A \cap B} S_j.
\end{equation}
The above implication stems from the fact that both products on the left-hand side can be multiplied by the Pauli strings in $A \cap B$.
Since these Pauli stings appear twice in each resulting product and they anticommute with any Pauli string different from themselves, they cancel out to identity incurring at most a change of sign.

Following a similar reasoning as above, if there are two distinct and disjoint subsets $A \subset \mathcal{S}'$ and $B \subset \mathcal{S}'$ such that $\prod_{S_i \in A} S_i \propto \prod_{S_j \in B} S_j$,
\begin{equation}
    \label{eq:irrelevant_intersection2}
    \prod_{S_i \in A \cup B} S_i \propto \mathbbm{1}_N,
\end{equation}
where $\mathbbm{1}_N$ is the identity operator in the Hilbert space of $N$ qubits.
In short, it is enough to prove that there is no subset $\mathcal{I} \subseteq \mathcal{S}' \subset \mathcal{S}$ fulfilling $\prod_{S_i \in \mathcal{I}} S_i \propto \mathbbm{1}_N$.
In what follows, we will prove this by showing that
\begin{equation}
    \label{eq:if_identity_then_not_subset}
    \prod_{S_i \in \mathcal{I}} S_i \propto \mathbbm{1}_N \Rightarrow \mathcal{I} = \mathcal{S},
\end{equation}
so no such $\mathcal{I} \subseteq \mathcal{S}'$ exists for any incomplete subset $\mathcal{S}'$ of $\mathcal{S}$.

Given a TT and a subset of its legs $\mathcal{I} \subseteq \mathcal{S}$, we can define a set of \textit{link multiplicities} $\{ \varphi_l \}$, where $\varphi_l$ is an integer defined for every link $l$ in the tree (be it an edge or a leg) and counts the number of paths from the root node to each of the legs in $\mathcal{I}$ that traverse link $l$.
Now, if we assume that $\prod_{S_i \in \mathcal{I}} S_i \propto \mathbbm{1}_N$, we can make the following observations:
\begin{itemize}
    \item[1.] For any node $u$ in the tree, the link multiplicities $\varphi_{l^{(u)}_x}$, $\varphi_{l^{(u)}_y}$, and $\varphi_{l^{(u)}_z}$ of the three links stemming downwards from $u$ must either be all even or all odd.
    This is a consequence of the fact that the product of Pauli strings in $\mathcal{I}$ results in a product of Pauli operators $X_u$, $Y_u$, and $Z_u$ on qubit $u$.
    Since these operators anticommute with one another, and their product must be proportional to identity according to our assumption above,
    \begin{equation}
        \left( X_u \right)^{\varphi_{l^{(u)}_x}} \left( Y_u \right)^{\varphi_{l^{(u)}_y}} \left( Z_u \right)^{\varphi_{l^{(u)}_z}} \propto \mathbbm{1},
    \end{equation}
    which can only be fulfilled if all three link multiplicities have equal parity.
    \item[2.] Consider a node $u$ different from the root, and let us refer to its upward edge multiplicity by $\varphi_{l^{(u)}_{\text{up}}}$.
    The downward link multiplicities are $\varphi_{l^{(u)}_x}$, $\varphi_{l^{(u)}_y}$, and $\varphi_{l^{(u)}_z}$, like above.
    If the assumption $\prod_{S_i \in \mathcal{I}} S_i \propto \mathbbm{1}_N$ holds, then $\varphi_{l^{(u)}_{\text{up}}}$ must have the same parity as the three downward links.
    This is a direct consequence of the fact that edge multiplicity is conserved,
    \begin{equation}
        \varphi_{l^{(u)}_{\text{up}}} = \varphi_{l^{(u)}_x} + \varphi_{l^{(u)}_y} + \varphi_{l^{(u)}_z},
    \end{equation}
    since every path that traverses $l^{(u)}_{\text{up}}$ must traverse one of the three downward links, and of observation 1.
    Indeed, the sum of an odd number of odd numbers is odd, and no odd number can be obtained by adding even numbers.
\end{itemize}

These two observations imply that the parity of the link multiplicities is conserved at each node, that is, all links reaching a node must have equal multiplicity parity if the assumption $\prod_{S_i \in \mathcal{I}} S_i \propto \mathbbm{1}_N$ is true.
Since the tree is connected, it follows that the multiplicity of all links in the graph must have the same parity.
Given that the legs in $\mathcal{I}$ have multiplicity one, all links in the graph must have odd multiplicity.
Thus, all legs in $\mathcal{S}$ must have multiplicity one and hence be in $\mathcal{I}$, which proves Eq.~\eqref{eq:if_identity_then_not_subset}.

\section{From Fock basis states to computational basis states}\label{app:prod_pres_proof}
In the main text, we showed that, with the pairing introduced in Sect.~\ref{sec:pairing}, the fermionic vacuum is mapped to $\ket{0}^{\otimes N}$, and that states of the form $a^{\dagger}_i \ket{\text{vac}_\text{f}}$ lead to computational basis states in qubit space.
We now show that this is also true for any Fock basis state.

Consider an arbitrary Fock basis state $\ket{\psi}$ in which the fermionic modes in the subset $\mathcal{F} \subseteq \{0, 1, \ldots, N-1 \}$ are occupied, that is, $\ket{\psi} = \prod_{k \in \mathcal{F}} a^{\dagger}_k \ket{\text{vac}_\text{f}}$.
Since all the creation operators in the expression are different, they anticommute, so $\ket{\psi}$ can be written, up to a sign, by applying them in an arbitrary order.

Recall that, given a TT mapping, every fermionic mode $j$ can be associated with a qubit $u_j$ according to the pairing strategy (see Eq.~\eqref{eq:pairing_identification}).
This identification allows us to associate an integer $h_j$ to every mode in $\mathcal{F}$ indicating how deep $u_j$ lies down the tree.
More precisely, $h_j$ is the topological distance between mode $j$'s associated qubit $u_j$ and the root node.
Now, consider a sequence $(R_0, \ldots, R_{\vert \mathcal{F} \vert - 1} )$ of the elements in $\mathcal{F}$ (that is, $R_i \in \mathcal{F}$ for all $i \in \{ 0, \ldots, \vert \mathcal{F} \vert - 1\}$ and $R_i = R_j \Leftrightarrow i = j$) following a top-down order, $h_{R_i} \leq h_{R_{i+1}}\, \forall i \in \{ 0, \ldots, \vert \mathcal{F} \vert - 1 \}$.
We can then construct $\ket{\psi}$ by applying the sequence of creation operators starting from the highest modes up in the tree, and following downwards,
\begin{equation}
	\ket{\psi} \propto \prod_{i=0, \ldots, \vert \mathcal{F} \vert - 1} a^{\dagger}_{R_i} \ket{\text{vac}_\text{f}}.
\end{equation}

Each fermionic operator $a^{\dagger}_j$ involves the two Majorana strings $S_{s_x^{(u_j)}}$ and $S_{s_y^{(u_j)}}$, which only differ on qubit $u_j$ (on which they act with $X_{u_j}$ and $Y_{u_j}$, respectively), and on all qubits in the $X$ and $Y$ branches lying below $u_j$ in the tree; each of the two Majorana strings acts with a $Z$ operator on the qubits on one of the branches, but trivially on the qubits in the other branch.
Therefore, if $\ket{\phi}$ is a computational basis state in which $u_j$ and all the qubits below it are in the $\ket{0}$ state, $(S_{s_x^{(u_j)}} - i S_{s_y^{(u_j)}}) \ket{\phi} = (S'_{s_x^{(u_j)}} - i S'_{s_y^{(u_j)}}) \ket{\phi}$, where the $Z$ Pauli operators on the qubits below $u_j$ have substituted with identities in the primed Pauli strings, as in Sect.~\ref{sec:pairing}.
Importantly, in the output vector, the state of $u_j$ and possibly of other qubits above $u_j$ in the tree are flipped, but not the state of qubits below $u_j$.
In addition, the vector remains a computational basis one.

With this setup, we can proceed in an inductive way.
First, it is clear from the above discussion (and the one in the main text) that the state $a^{\dagger}_{R_0} \ket{\text{vac}_\text{f}}$ in qubit space, let us denote it by $\ket{\phi_0}$, is a computational basis state.
Second, it can be seen that if the mapped state $\prod_{i=0, \ldots, n} a^{\dagger}_{R_i} \ket{\text{vac}_\text{f}}$, $\ket{\phi_n}$, is a computational basis state, then so is $\ket{\phi_{n+1}}$.
This follows from
\begin{equation}
	\ket{\phi_{n + 1}} = \frac{1}{2} \left( S_{s_x^{(u_{N + 1})}} - i S_{s_y^{(u_{n + 1})}} \right) \ket{\phi_n},
\end{equation}
where $u_{n + 1}$ is the qubit associated with the fermionic mode $R_{n + 1}$.
Since all the Majorana strings that must be applied to prepare $\ket{\phi_n}$ from $\ket{0}^{\otimes N}$ act on $u_{n + 1}$ and all qubits below it with either identity or with $Z$ (again, given that $h_{R_i} \leq h_{R_{n+1}}\, \forall i \in \{ 0, \ldots, n \}$), the state of each of those qubits must be $\ket{0}$.
As we have shown above, along with the condition that $\ket{\phi_n}$ be a computational basis state (which here is true by assumption) guarantees that $\ket{\phi_{n + 1}}$ is a computational basis state as well.

\newpage
\onecolumngrid
\section{Algorithms and routines in more detail}
\label{app:algorithms}
\LinesNumbered
\begin{algorithm*}[h]
    \caption{Qubit spanning tree subroutine}
    \label{alg:spanning_tree}
    
    Define physical connectivity graph $\mathcal{P}=(V_\mathcal{P}, E_\mathcal{P})$.

    Determine root node $r = \text{argmin}_u \max_v d(u, v; \mathcal{P})$, where $d(u, v; \mathcal{P})$ is the topological distance between $u$ and $v$ in $\mathcal{P}$.
    
    Define initial layer, $\mathcal{L}_0 = {r}$, height $h = 0$, and tree $\mathcal{T} = (V_\mathcal{T}, E_\mathcal{T})$ with $V_\mathcal{T} = E_\mathcal{T} = \emptyset$.
    \BlankLine

    \While{$\mathcal{L}_{h} \neq \emptyset$}{
    
        Define $\mathcal{L}_{h + 1} = \emptyset$.
        
        \For{$v \in \mathcal{L}_{h}$}{

            Define the set of unassigned neighbours of $v$, $\mathcal{N}_v = \{ w \in V_\mathcal{P} : (v, w) \in E_\mathcal{P} \wedge w \notin V_\mathcal{T} \}$.

            \If{$\abs{\mathcal{N}_v} > 3$}{
            
                Define $\mathcal{N}'_v \subset \mathcal{N}_v$ containing three nodes randomly chosen from $\mathcal{N}_v$.
                
                Set $\mathcal{N}'_v \rightarrow \mathcal{N}_v$.
                
            }

            Set $\mathcal{L}_{h + 1} \cup \mathcal{N}_v \rightarrow \mathcal{L}_{h + 1}$.
            
            Set $V_\mathcal{T} \cup \mathcal{N}_v \rightarrow V_\mathcal{T}$.

            Set $E_\mathcal{T} \cup \bigcup_{w \in \mathcal{N}_v} \{ (v, w), (w, v) \} \rightarrow E_\mathcal{T}$.
            
        }
    Set $h+1 \rightarrow h$.
    }    
    \BlankLine
    
    \For{$u \in V_\mathcal{P} \setminus V_\mathcal{T}$}{
        
        Determine set $\mathcal{A} \subseteq V_\mathcal{T} $ of nodes in $\mathcal{T}$ available to connect, $\mathcal{A} = \{ u \in V_\mathcal{T} : \abs{\{ v \in V_\mathcal{T} : (u, v) \in E_\mathcal{T} \}} + \delta_{u, r} < 4 \}$.
        
        Find set $\mathcal{C} \subseteq \mathcal{A}$ of closest nodes to $u$, $\mathcal{C} = \{ v \in \mathcal{A} : d(u, v; \mathcal{P}) = \min_w (\{ d(u, w; \mathcal{P}) : w \in \mathcal{A} \}) \}$.

        \If{$\abs{\mathcal{C}} > 1$}{
        
            Define $\mathcal{C}' \subset \mathcal{C}$ containing one node randomly chosen from $\mathcal{C}$.
            
            Set $\mathcal{C}' \rightarrow \mathcal{C}$.
            
        }
        
        Set $V_\mathcal{T} \cup \mathcal{C} \rightarrow V_\mathcal{T}$.

        Set $E_\mathcal{T} \cup \{ (u, v), (v, u) \} \rightarrow E_\mathcal{T}$, where $v \in \mathcal{C}$.
        
    }
    \BlankLine
    \KwRet{$\mathcal{T}$}
\end{algorithm*}

\begin{algorithm*}[h]
    \caption{Labelling subroutine}
    \label{alg:branching_sub}
    \SetKwProg{Fn}{Function}{:}{}
    \SetKwProg{Pr}{Procedure}{:}{}

    \Pr{Minimise delocalisation by maximising all-$Z$ branch length}{
        Find the longest path $\ell$ in $\mathcal{T}$.
        
        Associate a $Z$ label to every edge along $\ell$.        
    }
    
    \BlankLine
    \Pr{Homogeneous localisation}{
        For every node in the tree, add labels to each of its unlabelled descending edges with priority 1) $X$, 2) $Y$, and 3) $Z$ (that is, single edges are labelled with $X$ and double edges with $XY$). 
        
        Add labels to all legs.
    }
    
    \BlankLine
    \Pr{Heterogeneous localisation}{
        For every node in the tree, add labels to each of its unlabelled descending edges with priority 1) $Z$ (if available), 2) $X$, and 3) $Y$ (that is, single edges are labelled with $Z$ and double edges with $ZX$). 
        
        Add labels to all legs.
    }
\end{algorithm*}

\newpage
\section{\label{app:HH mode ops}The heavy-hexagon mappings explicitly}

\begin{table*}[h]
    \caption{\label{tab:hh_mode_ops} Qubit mode operators generated from the trees in Fig.~\ref{fig:mode_op_tree_folded}.
    Localised operators are ones which ones with raising/lowering $P^\pm$ operators acting on the $i$-th qubit.
    Specific delocalisation is clear from the number of Pauli $Z$ operators in the brackets.}
    \begin{ruledtabular}
        \begin{tabular}{ccc}
         &\multicolumn{2}{c}{$a_j^{(\dagger)}$}\\
         j & heterogeneous localisation  & heterogeneous localisation  \\ \hline
        0  &  $\frac{1}{2}(X_{0}Z_{1} \mp iY_{0}Z_{2})                                 $  &  $\frac{1}{2}(X_{0}Z_{1}Z_{4}Z_{8}Z_{14}Z_{22}Z_{31}Z_{34} \mp iY_{0}Z_{2}Z_{5}Z_{10}Z_{16}Z_{26}Z_{32}Z{35})  $ \\ 
        1  &  $\frac{1}{2} X_{0}(X_{1}Z_{4} \mp iY_{0})                                $  &  $X_{0}P^\pm_1                                                                                             $ \\ 
        2  &  $\frac{1}{2} Y_{0}(X_{2}Z_{5} \mp iY_{2})                                $  &  $Y_{0}P^\pm_2                                                                                             $ \\ 
        3  &  $Z_{0}P^\pm_3                                                       $       &  $Z_{0}P^\pm_3                                                                                             $ \\ 
        4  &  $\frac{1}{2} X_{0}X_{1}(X_{4}Z_{7} \mp iY_{4}Z_{8})                      $  &  $\frac{1}{2}X_{0} Z_{1} (X_{4}Z_{7}Z_{13}Z_{20} \mp Y_{4}  )                                                              $ \\ 
        5  &  $\frac{1}{2} Y_{0}X_{2}(X_{5}Z_{9} \mp iY_{5}Z_{10})                     $  &  $\frac{1}{2} Y_{0} Z_{2} (X_{5}Z_{9}Z_{15}Z_{24} \mp Y_{5}  )                                                              $ \\ 
        6  &  $\frac{1}{2} Z_{0}Z_{3}(X_{6}Z_{11} \mp iY_{6})                          $  &  $\frac{1}{2} Z_{0} Z_{3} (X_{6}Z_{11}Z_{17}Z_{28} \mp Y_{6}  )                                                             $ \\ 
        7  &  $\frac{1}{2} X_{0}X_{1}X_{4}(X_{7}Z_{13} \mp iY_{7})                     $  &  $X_{0} Z_{1} X_{4} P^\pm_7                                                                                $ \\ 
        8  &  $\frac{1}{2} X_{0}X_{1}Y_{4}(X_{8}Z_{14} \mp iY_{8})                     $  &  $X_{0} Z_{1} Z_{4} P^\pm_8                                                                                $ \\ 
        9  &  $\frac{1}{2} Y_{0}X_{2}X_{5}(X_{9}Z_{15} \mp iY_{9})                     $  &  $Y_{0} Z_{2} X_{5} P^\pm_9                                                                                $ \\ 
        10 &  $\frac{1}{2} Y_{0}X_{2}Y_{5}(X_{10}Z_{16} \mp iY_{10})                   $  &  $Y_{0} Z_{2} Z_{5} P^\pm_{10}                                                                             $ \\ 
        11 &  $\frac{1}{2} Z_{0}Z_{3}Z_{6}(X_{11}Z_{17} \mp iY_{11})                   $  &  $Z_{0} Z_{3} X_{6} P^\pm_{11}                                                                             $ \\ 
        12 &  $Z_{0}Z_{3}Z_{6} P^\pm_{12}                                           $     &  $Z_{0} Z_{3} Z_{6} P^\pm_{12}                                                                             $ \\ 
        13 &  $\frac{1}{2} X_{0}X_{1}X_{4}X_{7}(X_{13}Z_{19} \mp iY_{13}Z_{20})        $  &  $\frac{1}{2}X_{0} Z_{1} X_{4} Z_{7}  (X_{13}Z_{19} \mp iY_{13})                                                           $ \\ 
        14 &  $\frac{1}{2} X_{0}X_{1}Y_{4}X_{8}(X_{14}Z_{21} \mp iY_{14}Z_{22})        $  &  $\frac{1}{2}X_{0} Z_{1} Z_{4} Z_{8}  (X_{14}Z_{21} \mp iY_{14})                                                           $ \\ 
        15 &  $\frac{1}{2} Y_{0}X_{2}X_{5}X_{9}(X_{15}Z_{23} \mp iY_{15}Z_{24})        $  &  $\frac{1}{2}Y_{0} Z_{2} X_{5} Z_{9}  (X_{15}Z_{23} \mp iY_{15})                                                           $ \\ 
        16 &  $\frac{1}{2} Y_{0}X_{2}Y_{5}X_{10}(X_{16}Z_{25} \mp iY_{16}Z_{26})       $  &  $\frac{1}{2} Y_{0} Z_{2} Z_{5} Z_{10} (X_{16}Z_{25} \mp iY_{16})                                                           $ \\ 
        17 &  $\frac{1}{2} Y_{0}X_{2}X_{5}X_{9}(X_{15}Z_{23} \mp iY_{15}Z_{24})        $  &  $\frac{1}{2} Z_{0} Z_{3} X_{6} Z_{11} (X_{17}Z_{27} \mp iY_{17})                                                           $ \\ 
        18 &  $\frac{1}{2} Y_{0}X_{2}Y_{5}X_{10}(X_{16}Z_{25} \mp iY_{16}Z_{26})       $  &  $\frac{1}{2} Z_{0} Z_{3} Z_{6} Z_{12} (X_{18}Z_{29} \mp iY_{18})                                                           $ \\ 
        19 &  $X_{0}X_{1}X_{4}X_{7}X_{13}P_{19}^\pm                                   $   &  $X_{0} Z_{1} X_{4} Z_{7} X_{13} P^\pm_{19}                                                                $ \\ 
        20 &  $X_{0}X_{1}X_{4}X_{7}Y_{13}P_{20}^\pm                                   $   &  $X_{0} Z_{1} X_{4} Z_{7} Z_{13} P^\pm_{20}                                                                $ \\ 
        21 &  $X_{0}X_{1}Y_{4}Y_{8}X_{14}P_{21}^\pm                                   $   &  $X_{0} Z_{1} Z_{4} Z_{8} X_{14} P^\pm_{21}                                                                $ \\ 
        22 &  $\frac{1}{2} X_{0}X_{1}Y_{4}X_{8}Y_{14}(X_{22}Z_{31} \mp iY_{22})        $  &  $X_{0} Z_{1} Z_{4} Z_{8} Z_{14} P^\pm_{22}                                                                $ \\ 
        23 &  $Y_{0}X_{2}X_{5}X_{9}X_{15}P_{23}^\pm                                   $   &  $Y_{0} Z_{2} X_{5} Z_{9} X_{15} P^\pm_{23}                                                                $ \\ 
        24 &  $Y_{0}X_{2}X_{5}X_{9}Y_{15}P_{24}^\pm                                   $   &  $Y_{0} Z_{2} X_{5} Z_{9} Z_{15} P^\pm_{24}                                                                $ \\ 
        25 &  $Y_{0}X_{2}Y_{5}X_{10}X_{16}P_{25}^\pm                                  $   &  $Y_{0} Z_{2} Z_{5} Z_{10} X_{16} P^\pm_{25}                                                               $ \\ 
        26 &  $\frac{1}{2} Y_{0}X_{2}Y_{5}X_{10}Y_{16}(X_{26}Z_{32} \mp iY_{26})       $  &  $Y_{0} Z_{2} Z_{5} Z_{10} Z_{16} P^\pm_{26}                                                               $ \\ 
        27 &  $Z_{0}Z_{6}X_{6}X_{11}X_{17}P_{27}^\pm                                  $   &  $Z_{0} Z_{3} X_{6} Z_{11} X_{17} P^\pm_{27}                                                               $ \\ 
        28 &  $Z_{0}Z_{6}X_{6}X_{11}Y_{17}P_{28}^\pm                                  $   &  $Z_{0} Z_{3} X_{6} Z_{11} Z_{17} P^\pm_{28}                                                               $ \\ 
        29 &  $Z_{0}Z_{6}X_{6}Z_{12}X_{18}P_{29}^\pm                                  $   &  $Z_{0} Z_{3} Z_{6} Z_{12} X_{18} P^\pm_{29}                                                               $ \\ 
        30 &  $ Z_{0}Z_{6}X_{6}Z_{12}Z_{18}P_{30}^\pm                  $                  &  $Z_{0} Z_{3} Z_{6} Z_{12} Z_{18} P^\pm_{30}                                                               $ \\ 
        31 &  $\frac{1}{2} X_{0}X_{1}Y_{4}Y_{8}Y_{14}X_{22}(X_{31}Z_{34} \mp iY_{31})  $  &  $X_{0} Z_{1} Z_{4} Z_{8} Z_{14} Z_{22} P_{31}^\pm                                                         $ \\ 
        32 &  $\frac{1}{2} Y_{0}X_{2}Y_{5}X_{10}Y_{16}X_{26}(X_{32}Z_{35} \mp iY_{32}) $  &  $Y_{0} Z_{2} Z_{5} Z_{10} Z_{16} Z_{26} P_{32}^\pm                                                        $ \\ 
        33 &  $ Z_{0}Z_{6}X_{6}Z_{12}Z_{18}Z_{30}P_{33}^\pm            $                  &  $Z_{0} Z_{3} Z_{6} Z_{12} Z_{18} Z_{30} P^\pm_{33}                                                        $ \\ 
        34 &  $X_{0}X_{1}Y_{4}X_{8}Y_{14}X_{22}X_{31}P_{34}^\pm                       $   &  $X_{0} Z_{1} Z_{4} Z_{8} Z_{14} Z_{22} Z_{31} P_{34}^\pm                                                  $ \\ 
        35 &  $Y_{0}X_{2}Y_{5}X_{10}Y_{16}X_{26}X_{32}P_{35}^\pm                      $   &  $Y_{0} Z_{2} Z_{5} Z_{10} Z_{16} Z_{26} Z_{32} P_{35}^\pm                                                 $ \\ 
        36 &  $ Z_{0}Z_{6}X_{6}Z_{12}Z_{18}Z_{30}Z_{33}P_{36}^\pm      $                  &  $Z_{0} Z_{3} Z_{6} Z_{12} Z_{18} Z_{30} Z_{33} P^\pm_{36}                                                 $ \\ 

    \end{tabular}
    \end{ruledtabular}
\end{table*}

\newpage

\twocolumngrid

\end{document}